\documentclass[aps,prb,article,twocolumn,preprintnumbers,amsmath,amssymb,superscriptaddress]{revtex4-2}
\usepackage[colorlinks=true, linkcolor=blue, urlcolor=blue,citecolor=blue]{hyperref}
\usepackage{physics}
\usepackage{graphicx} 
\usepackage{dcolumn}
\usepackage{bm}
\usepackage{comment}
\usepackage{amssymb}
\usepackage{amsmath}
\usepackage{mathrsfs}
\usepackage{epigraph}
\usepackage{braket}
\usepackage{gensymb}
\usepackage{lmodern}
\usepackage{lipsum}
\usepackage{marvosym}
\usepackage{ tipa }
\usepackage{bbold}
\usepackage{dsfont}
\usepackage{esint}
\usepackage{mathdots}
\usepackage{xcolor}
\usepackage{appendix}
\usepackage{natbib}
\usepackage{multirow}
\usepackage{float}
\usepackage{epstopdf}

\begin{document}
\title{Charge-Density Wave in Extremely Overdoped Cuprates Driven by Phonons} 	
\author{Jiarui Liu}
\affiliation{Department of Physics and Astronomy, Clemson University, Clemson, SC 29631, United States}
\author{Shaozhi Li}
\affiliation{Department of Physics and Astronomy, Clemson University, Clemson, SC 29631, United States}
\author{Edwin W. Huang}
\affiliation{Department of Physics and Astronomy, University of Notre Dame, Notre Dame, IN 46556, United States}
\affiliation{Stavropoulos Center for Complex Quantum Matter, University of Notre Dame, Notre Dame, IN 46556, United States}
\author{Yao Wang}
\email{yao.wang@emory.edu}
\affiliation{Department of Physics and Astronomy, Clemson University, Clemson, SC 29631, United States}
\affiliation{Department of Chemistry, Emory University, Atlanta, GA 30322, United States}
\date{\today}
\begin{abstract}
  Recent resonant x-ray scattering (RXS) experiments revealed a novel charge order in extremely overdoped La$_{2-x}$Sr$_x$CuO$_4$ (LSCO) [{Phys. Rev. Lett.}~\textbf{131},116002]. The observed charge order appears around the $(\pi/3,0)$ wavevector, distinct from the well-known stripe fluctuations near 1/8 doping, and persists from cryogenic temperatures to room temperature. To investigate the origin of this charge order in the overdoped regime, we use determinant quantum Monte Carlo (DQMC) simulations to examine correlated models with various interactions. We demonstrate that this distinctive CDW originates from remnant correlations in extremely overdoped cuprates, with its specific pattern shaped by interactions beyond the Hubbard model, particularly electron-phonon couplings. The persistence of the $(\pi/3,0)$ wavevector across different doping levels indicates the presence of nonlocal couplings. Our study reveals the significant role of phonons in cuprates, which assist correlated electrons in the formation of unconventional phases.
\end{abstract}
\maketitle
\section{Introduction}

Unconventional superconductivity (SC) in cuprates has attracted extensive experimental and theoretical studies\,\cite{bednorz1986possible, dagotto1994correlated,lee2006doping,damascelli2003angle,agterberg2020physics}. In addition to its promising applications in energy and quantum technology, the exploration of cuprates has been driven by numerous complex phases of cuprates\,\cite{keimer2015quantum}. These phases can coexist with or compete against SC, challenging conventional solid-state theories. A particularly significant phase among these is the charge density wave (CDW). In conventional BCS superconductors, phonons mediate an effective electron-electron attraction that gives rise to both mobile Cooper pairs and immobile charge modulations. These two states compete against each other as evidenced by investigations into Holstein-like models\,\cite{li2017competing,shi2018variational,esterlis2018breakdown}. In cuprates, despite having distinct pairing symmetry and potential differences in the pairing mechanisms from BCS theory, many experiments have revealed the presence of CDW orders or fluctuations near the SC phase. Their proximity suggests an intertwined origin of CDW and SC, even in the unconventional and high-$T_c$ materials\,\cite{davis2013concepts,peli2017mottness,seibold2021strange,lee2022generic,kang2019evolution}.

\begin{figure}[!b]
\begin{center}
\includegraphics[width=9cm]{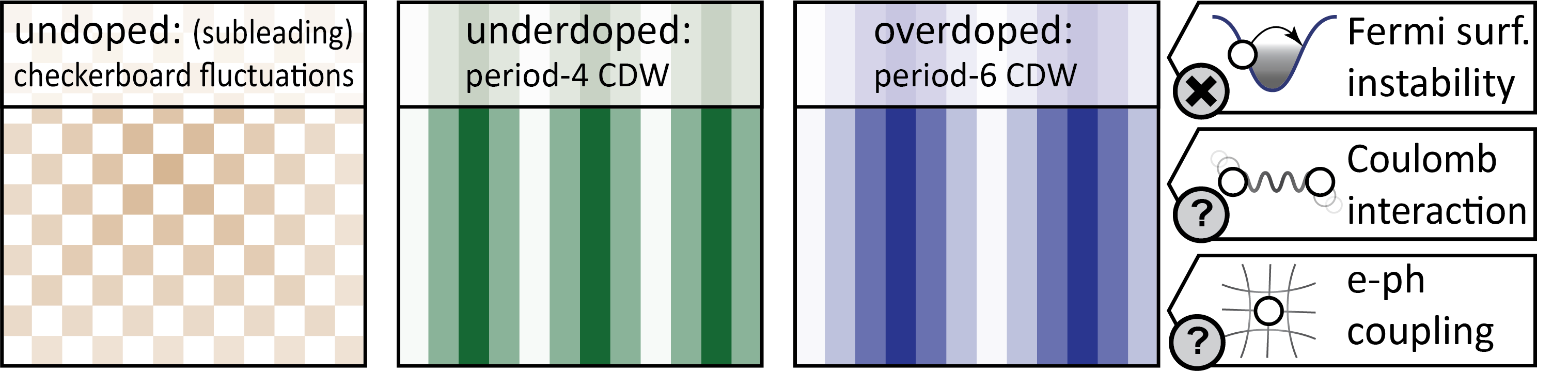}\vspace{-2mm}
\caption{Schematic illustrating different types of charge fluctuations or order observed from experiments across various doping regimes of cuprates. The origin of the overdoped period-6 CDW serves as the primary focus of this work.\vspace{-5mm}
}
\label{figcartoon}
\end{center}
\end{figure}

Research on CDW has predominantly focused on the underdoped and optimally doped cuprates\,\cite{tranquada1995evidence,comin2016resonant,hucker2011stripe,campi2015inhomogeneity}, highlighting the interplay between SC, the pseudogap phase, and CDW\,\cite{da2014ubiquitous,fradkin2015colloquium,loret2019intimate}. Particularly, near 12.5\% hole doping, a charge order has been detected that manifests as a unidirectional stripe behavior with a periodicity of 4 unit cells [see Fig.~\ref{figcartoon}]. Advanced numerical many-body methods have successfully described this stripe order in the context of the Hubbard and Hubbard-Holstein model\,\cite{huang2018stripe, zheng2017stripe,karakuzu2022stripe}. Remarkably, the CDW emerging in these simulations at around 12.5\% hole doping exhibits a notable competition with $d$-wave superconductivity\,\cite{jiang2021groundstate,jiang2019superconductivity,xu2023coexistence}. 

Recent observations of CDWs in overdoped cuprates, specifically in (Bi,Pb)$_{2.12}$Sr$_{1. 88}$CuO$_{6+\delta}$ (Bi2201)\,\cite{peng2018re}, have sparked a renewed interest in understanding their role in high-$T_c$ superconductors. This CDW phenomenon was subsequently observed in extremely overdoped LSCO, in the form of a charge modulation of extending over 6 unit cells in the antinodal direction [see Fig.~\ref{figcartoon}] and persists up to room temperature\,\cite{li2023prevailing}. (Here, we refer to the $(H,0)$ and $(H,H)$ directions as ``antinodal'' and ``nodal'', respectively.) This CDW starts to develop at 35\% doping and maximizes in intensity at approximately 50\% doping. Despite the expected screening of Coulomb interactions in the overdoped regime, Fermi-surface instabilities cannot explain the origin of this CDW\,\cite{li2023prevailing}. This suggests that the CDW is likely driven by other interactions. Importantly, this overdoped regime, being distanced from the SC and pseudogap phases, may be less affected by the dominant spin fluctuations, providing a unique opportunity to explore intrinsic yet subleading interactions in cuprates.

For this purpose, we examine the extremely overdoped cuprates using various models, including the Hubbard model, the Hubbard-Holstein model, and their variants. Our findings reveal that while the Hubbard interaction correctly produces a correlation-induced charge instability peaking at 50\% doping, it fails to capture the charge pattern with wavevector $(\pi/3,0)$. The inclusion of electron-phonon coupling (EPC), particularly nonlocal EPC, shifts the ordering wavevector from $(\pi,\pi)$ to the experimentally observed $(\pi/3,0)$. Building upon this framework, we quantify the strength and distribution of the EPC, addressing the experimental findings in LSCO. This work highlights the essential contributions of EPC to correlated phases in cuprates.

The organization of this paper is as follows. We first introduce the models and the DQMC method in Sec.~\ref{sec:model}. Next, we investigate charge susceptibilities in the extremely overdoped Hubbard model, demonstrating the absence of the expected CDW in Sec.~\ref{sec:Hubbard}. We then systematically examine the impact of phonons, with local and nonlocal EPCs, in Sec.~\ref{sec:phonon}. This analysis is further generalized to bond phonons in Sec.~\ref{sec:nonHolstein}. Finally, Sec.~\ref{sec:conclusion} summarizes the conclusion and discusses potential connection to superconductivity in overdoped cuprates.

\section{Models and Method}\label{sec:model}

We begin our theoretical analysis by considering the single-band Hubbard model\,\cite{hubbard1963electron,zhang1988effective}:
\begin{eqnarray}\label{eq:Hubbard}
\mkern-2mu\mathcal{H}_{\rm Hubbard}\mkern-2mu=-\mkern-2mu\sum_{ij\sigma}\mkern-2mut_{ij} c_{i\sigma}^\dagger c_{j\sigma}-\mu\mkern-2mu\sum_{i\sigma} n_{i\sigma}+\mkern-2mu\sum_{i}U n_{i\uparrow}n_{i\downarrow}\mkern-2mu\,,
\end{eqnarray}
where $c_{i\sigma}$ ($c_{i\sigma}^\dagger$) annihilates (creates) an electron at site $i$ with spin $\sigma$ and $n_{i\sigma}=c_{i\sigma}^\dagger c_{i\sigma}$ is the corresponding density operator. The hopping term is defined by the one-electron integral $t_{ij}$ between Wannier wavefunctions at sites $i$ and $j$. Under the tight-binding approximation, we limit the hopping to the nearest and next-nearest neighbors, denoted as $t$ and $t^{\prime}$, respectively. This effective model for cuprates simplifies the electronic Coulomb interaction to the on-site Hubbard $U$ term and the average doping to a chemical potential ($\mu$) shift. To reflect the LSCO electronic structure determined by first-principles simulations and ARPES fitting\,\cite{pavarini2001band,yoshida2006systematic}, we use $t=250$\,meV, $t^{\prime}=-0.15t$, and $U=8t$ here. Notably, the experimental benchmarks for these parameters were obtained for dopings below 30\%. Due to the lack of ARPES experiments for the extremely overdoped regime, we assume the parameters in Eq.~\eqref{eq:Hubbard} remain consistent across all doping levels. This assumption is supported by the resilience of the simulated charge susceptibilities to variations in band parameters, as discussed in Secs.~\ref{sec:Hubbard} and \ref{sec:phonon}. 

The major simulations in this paper extend the Hubbard models by incorporating additional EPCs. We primarily focus on site phonons that couple to the electron density ($n_{i\sigma}$) due to their direct impact on charge. When the interaction is local, the resulting model corresponds to the Hubbard-Holstein model, described by the Hamiltonian:
\begin{eqnarray}\label{eq:HubbardHolstein}
        \mathcal{H}_{\rm HH}\mkern-2mu= \mkern-2mu\mathcal{H}_{\rm Hubbard}\mkern-2mu+\mkern-4mu \sum_i\mkern-4mu\left[ \mkern-1mu\frac{M}{2}\omega_{\rm ph}^2 \mkern-1muX_i^2\mkern-2mu+\mkern-2mu\frac{P_i^2}{2M}\mkern-1mu\right]\mkern-2mu-\mkern-4mu\sum_{i\sigma}\mkern-2mug X_i n_{i\sigma}\,.\hspace{0.5cm}
 \end{eqnarray}
Here, ${X}_i$ (${P}_i$) denotes the lattice displacement (momentum) at lattice site $i$, $g$ is the onsite EPC strength, $M$ is the phonon oscillator mass (set as $1t^{-1}$), and $\omega_{\rm ph}$ is the phonon frequency. In the latter sections of this paper, we also consider non-local EPC, including the nearest-neighbor coupling $g^\prime$, next-nearest-neighbor coupling $g^{\prime\prime}$, next-next-nearest-neighbor coupling $g^{\prime\prime\prime}$, and 4th-nearest-neighbor coupling $g^{\prime\prime\prime\prime}$. These models, with these nonlocal EPCs, are collectively referred to as the Hubbard-extended-Holstein (HEH) model:
\begin{eqnarray}\label{eq:HubbardExtendedHolstein}
        \mathcal{H}_{\rm HEH}&=& \mathcal{H}_{\rm HH}\mkern-2mu-\mkern-2mu\sum_{i,\sigma}\mkern-2mu\bigg(\mkern-1mu\sum_{\langle i,j\rangle}\mkern-2mug^\prime  \mkern-2muX_{i} n_{j\sigma}\mkern-2mu-\mkern-4mu\sum_{\langle \mkern-3mu\langle i,j\rangle\mkern-3mu\rangle}\mkern-2mug^{\prime\prime} X_{i} n_{j\sigma}\nonumber\\
        &&-\mkern-4mu\sum_{\langle \mkern-3mu\langle\mkern-3mu\langle i,j\rangle\mkern-3mu\rangle\mkern-3mu\rangle}\mkern-2mug^{\prime\prime\prime} X_{i} n_{j\sigma}-\mkern-4mu\sum_{\langle \mkern-3mu\langle\mkern-3mu\langle\mkern-3mu\langle i,j\rangle\mkern-3mu\rangle\mkern-3mu\rangle\mkern-3mu\rangle}\mkern-2mug^{\prime\prime\prime\prime} X_{i} n_{j\sigma}\bigg)        \,.
\end{eqnarray}
Through with various EPCs, it is important to note the on-site Coulomb interaction $U$ remains the dominant interaction in all above models, reinforcing the strongly correlated nature of cuprates. As discussed in Sec.~\ref{sec:Hubbard}, this strong correlation is crucial for accurately reproducing the doping dependence observed in experiments and linking it to the well-known 1/8-doped stripes\,\cite{tranquada1995evidence,huang2017numerical}. 

We employ the determinant quantum Monte Carlo (DQMC) algorithm to simulate these strongly correlated models, considering the persistence of the overdoped CDW at high temperatures\,\cite{peng2018re, li2023prevailing}. DQMC is an unbiased quantum many-body method, which transforms the thermal density matrix into a summation over Hubbard–Stratonovich field configurations, which are then estimated through stochastic importance sampling\,\cite{blankenbecler1981monte,white1989numerical}. Our primary goal is to address the charge order observed in the RXS experiments, with a focus on the charge susceptibility $\chi_{\rm c}(\mathbf{q},\omega)$ at $\omega=0$ 
\begin{equation}
\begin{split}
\chi_{\rm c}(\mathbf{q},\omega\mkern-2mu=\mkern-2mu0)=&\int_0^\beta d\tau \sum_{i,j} e^{-i\mathbf{q}\cdot(\mathbf{r_i}-\mathbf{r_j})}\\ 
&\times\big[\braket{n_i(\tau) n_j(0)}-\braket{n_i(\tau)}\braket{n_j(0)} \big]\,.
\end{split}
\end{equation}
Here, $\beta=1/T$ denotes the inverse temperature, and $n_{i}=n_{i,\uparrow} + n_{i,\downarrow}$ represents the electron density at site $i$. If interactions are neglected, $\chi_{\rm c}$ can be evaluated using the Lindhard response function. However, as discussed in Ref.~[\onlinecite{li2023prevailing}], this estimation fails to explain the experimentally observed CDW. 

\begin{figure*}[!t]
\begin{center}
\includegraphics[width=18.2cm]{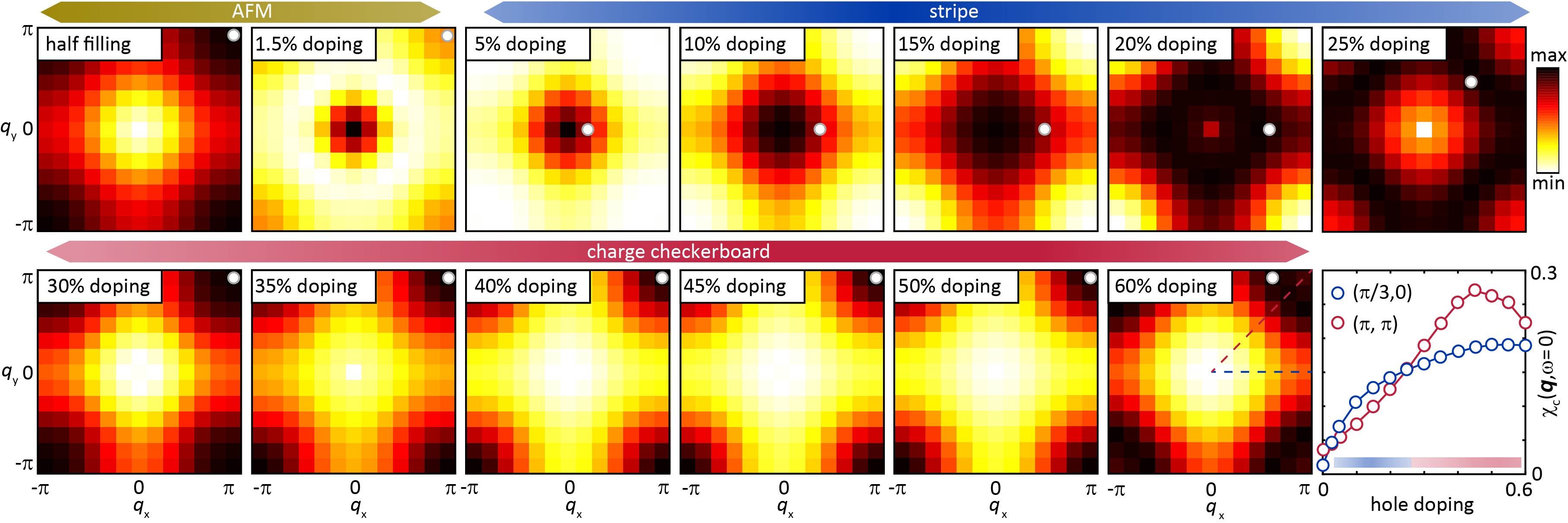}\vspace{-4mm}
\caption{\label{fig:Hubbard}
The charge susceptibility $\chi_{\rm c}(\mathbf{q},\omega=0)$ obtained from the Hubbard model on a $12\times 12 $ square lattice at temperature $T=0.4t$. The susceptibility is normalized for each doping (panel). For doping level 5\%-20\%, the white circles mark the wavevector fitted by four symmetric peaks; for other doping levels, they mark the maximal instability among all momenta. The upper ribbons indicate the corresponding phase. The last panel summarizes the doping dependence of charge susceptibility at $\mathbf{q}=(\pi,\pi)$ and experimentally relevant $(\pi/3,0)$.\vspace{-3mm}
 }
\end{center}
\end{figure*}

Simulating the Hubbard model with non-local site phonons presents significant challenges compared to the Hubbard and Hubbard-Holstein models. The latter two models with only local interactions require rank-1 DQMC updates; in contrast, the presence of nonlocal EPCs affect multiple electron sites at once, necessitating a rank-$r$ update. Here, $r$ denotes the number of electron sites coupled to a single phonon. For example, the HEH model involving $g^\prime $ and $g^{\prime\prime}$ requires rank-9 updates (\textit{e.g.}, 1 + 4 + 4). The DQMC update is realized by the sampling weight ratio between two configurations, with the electronic part given by:
\begin{equation}
  \begin{split}
    R^\sigma \mkern-2mu&=\det[I+\Delta(\tau)(I-G^\sigma(\tau,\tau)]\\
    &=\det\mkern-2mu\left(\mkern-2mu\begin{array}{cccc}
         1\mkern-2mu+\mkern-2mu\Delta_{i_1} \mkern-2muA_{i_1,i_1} & \Delta_{i_1} \mkern-2muA_{i_1,i_2} & \cdots &\Delta_{i_1} \mkern-2muA_{i_1,i_r}\\
      \Delta_{i_2}\mkern-2mu A_{i_2,i_1} & \ddots & \ddots & \vdots\\
      \vdots& \ddots& \ddots& \vdots\\
      \Delta_{i_r} \mkern-2muA_{i_r,i_1}& \cdots & \cdots& 1\mkern-2mu+\mkern-2mu\Delta_{i_r} \mkern-2muA_{i_r,i_r}
    \end{array}\mkern-2mu\right)\,
  \end{split}.
\end{equation}
In this equation, the $\Delta(\tau)$ matrix contains $r$ non-zero elements, corresponding to the rank-$r$ update\,\cite{gubernatis2016quantum}. The phonon energy contribution, $e^{\int_0^\beta d\tau E(\tau)}$ is also included in the total partition function. Thus, the update ratio for the Metropolis algorithm follows
\begin{equation}
  R_{\rm tot}=R^\uparrow R^\downarrow \exp(-\delta\tau \Delta E).\hspace{0.5cm}
\end{equation}
Here, $\Delta E(x_1,\cdots,x_\tau,\cdots)= E(x_1,\cdots,x_\tau+\delta x,\cdots)-E(x_1,\cdots,x_\tau,\cdots)$ represents the energy difference between two phonon configurations, with $x_\tau$ being the phonon displacement at imaginary time $\tau$.

Besides the HEH model with site phonons and the DQMC method introduced so far, the subsequent sections will also employ a zero-temperature method for benchmarking (in Sec.~\ref{sec:phonon:temperature}) and discuss the effects induced by bond phonons (in Sec.~\ref{sec:nonHolstein}). To maintain the clarity of the discussion and avoid distracting the reader from the primary focus, we have postponed the introduction of these supplementary models and methods until their respective sections.

\section{Absence of Overdoped CDW in the Hubbard Model}\label{sec:Hubbard}
Using the Hubbard model, we examine the influence of strong electron-electron correlations on the behavior of extremely overdoped cuprates. Figure \ref{fig:Hubbard} presents the doping dependence and momentum dependence of $\chi_{\rm c}$ at $T=0.4t$, corresponding to 1000\,K. A $12\times 12$ square cluster is employed to maintain $D_{4h}$ symmetry, ensuring an unbiased comparison of charge instabilities along the nodal and antinodal directions. Given the substantial variation in charge susceptibility --- ranging from nearly zero at half-filling to significant levels at high doping --- we normalize $\chi_{\rm c}(\mathbf{q},\omega)$ by its maximum intensity at each doping level to highlight the relative spectral weight distribution across momentum space.
 
As the doping increases, the Hubbard model reveals three distinct behaviors in charge fluctuations. At and near half-filling, the system is dominated by an insulating antiferromagnetic (AFM) order, which suppresses charge fluctuations, resulting in a smaller $\chi_{\rm c}$. The remaining charge fluctuations are centered  around the wavevector $(\pi,\pi)$, corresponding to the subdominant doublon-hole fluctuations between nearest neighbors\,\cite{nowadnick2012competition}. At approximately $5\%$ doping, the system transitions from an AFM state to stripe fluctuations, evidenced by the emergence of prevailing wavevectors along the antinodal direction. In finite systems without $C_4$ symmetry breaking, the peaks in charge susceptibility manifest in both the $x$ and $y$ directions (see Appendix \ref{app:Hubbard})\,\cite{huang2017numerical, huang2018stripe}.

Beyond 25\% hole doping, the stripe fluctuations are gradually replaced by checkerboard charge fluctuations with $\mathbf{q}=(\pi,\pi)$. These (spatially) short-range fluctuations stems from strong correlations caused by the repulsive Hubbard $U$, which tends to associate one hole with a neighboring doublon. As the doping increases from 25\% to 50\%, the momentum distribution of charge fluctuations remains qualitatively consistent, while the overall intensity increases rapidly. This rise in intensity is attributed to unraveled charge carriers from the AFM background. However, this upward trend halts, and the susceptibility starts to drop at around quarter filling (50\% doping), where a singly-occupied checkerboard pattern develops. Further doping beyond 50\% disrupts this checkerboard pattern, leading to a reduction in intensity. 

The doping dependence of the charge susceptibility, obtained from the Hubbard model, successfully captures the experimental observations of the overall intensity in extremely overdoped cuprates\,\cite{li2023prevailing}. The maximum at quarter filling reflects the tendency towards a singly-occupied checkerboard pattern, as the Hubbard interaction suppresses double occupancy. Consequently, the experimentally observed doping dependence reflects the remnant correlations present in doped cuprates\,\cite{dean2013persistence, jia2014persistent, graf2007universal,le2011intense,lee2014asymmetry,ishii2014high}, which cannot be explained by the Fermi-surface instability alone [see discussions in Appendix \ref{app:Hubbard} and in Ref.~[\onlinecite{li2023prevailing}]. 

\begin{figure}[!t]
\begin{center}
\includegraphics[width=\columnwidth]{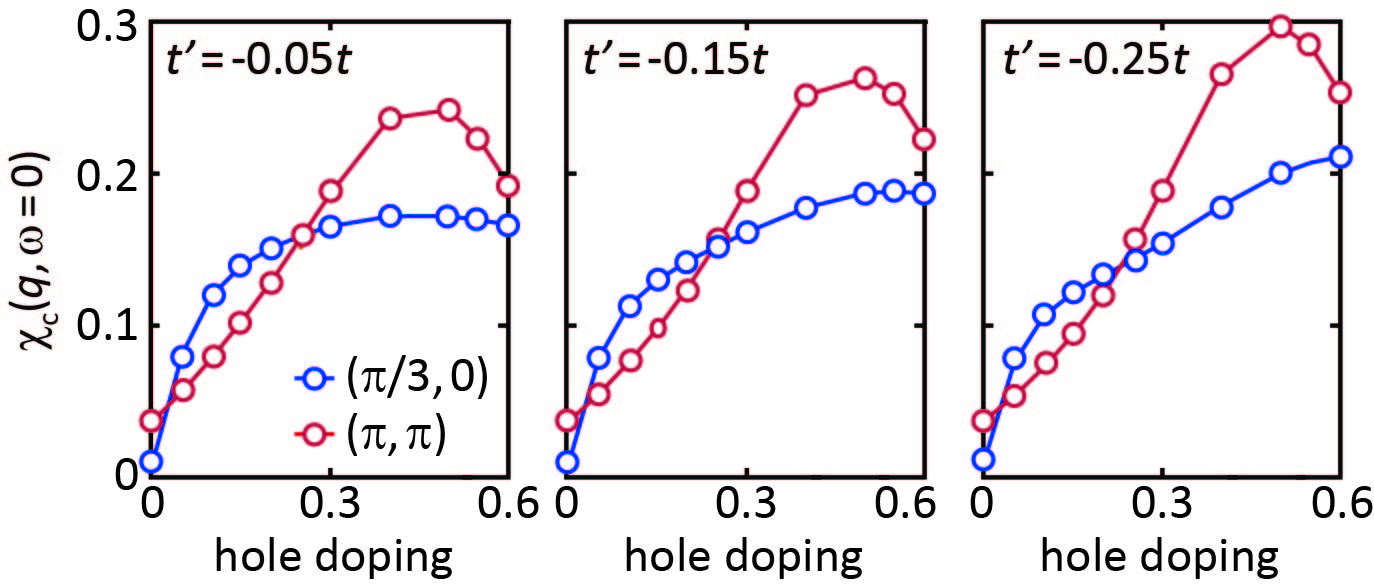}\vspace{-3mm}
\caption{\label{fig:HubbardtprimeDoping}
    The doping dependence of charge susceptibilities at $(\pi,\pi)$ and $(\pi/3,0)$, obtained from the Hubbard model with $U=8t$ and various $t^\prime$ values. All simulations are performed at $T=0.4t$. The error bars are smaller than the symbol sizes.\vspace{-8mm}
}
\end{center}
\end{figure}

To further investigate the influence of Fermi surface and band structure, we examine the $t^\prime$ dependence of the charge susceptibility across varying levels of hole doping. As shown in Fig.~\ref{fig:HubbardtprimeDoping}, the charge susceptibilities at $(\pi,\pi)$ and $(\pi/3,0)$ exhibit similar doping dependence across different $t^\prime$ values, representing specific band structures and materials. The results consistently show that charge susceptibility at $(\pi,\pi)$ exceeds that at $(\pi/3,0)$ in the heavily doped regime. Thus, the emergence of a checkerboard charge instability at quarter filling, rather than a $(\pi/3,0)$ charge order, is independent of specific band structures in the Hubbard model, although the exact intensities vary quantitatively.

Our simulations using the Hubbard model with various parameters demonstrate that, although the model accurately captures the doping-dependent behavior driven by correlations, it fails to precisely replicate the momentum distribution seen in experiments\,\cite{li2023prevailing}. Once the Hubbard model is doped beyond the stripe regime, the charge response consistently manifests dominance at $\mathbf{q}=(\pi,\pi)$ across all band structures relevant for cuprates. Thus, we conclude that the Hubbard model is \textit{insufficient} for explaining the experimentally observed overdoped CDW. 

\begin{figure*}[!t]
\begin{center}
\includegraphics[width=18.2cm]{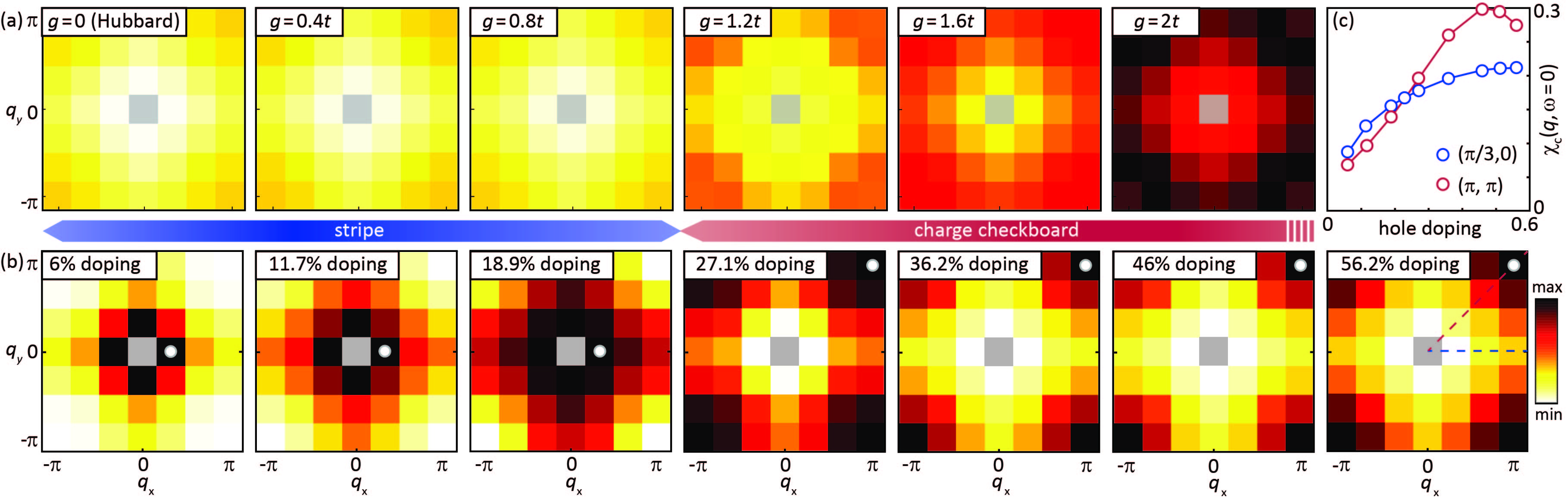}\vspace{-4mm}
\caption{\label{fig:HubbardHolstein}
(a) Charge susceptibility as a function of EPC strength at various momenta, simulated using the quarterly filled Hubbard-Holstein model with $U=8t$ at $T=0.4t$. (b) Doping dependence of charge susceptibilities, simulated using the Hubbard-Holstein model with the EPC $g=t$. The colormap in (a) is fixed to highlight intensity variations, while it is normalized within each panel of (b) to highlight the relative distribution. (c) Charge susceptibility at the dominant $(\pi,\pi)$ and the experimentally relevant $(\pi/3,0)$.
}

\end{center}
\end{figure*}

\section{Phonon-Mediated Overdoped CDW}\label{sec:phonon}
A natural extension of the Hubbard model involves incorporating additional interactions. Recent investigations into 1D cuprate chains have revealed the significant role of attractive nonlocal interactions mediated by phonons\,\cite{chen2021anomalously, wang2021phonon, tang2023traces, feiguin2023effective, wang2022spectral, banerjee2023ground, shen2024signatures}. These interactions are essential for accurately describing the behavior of cuprates, especially superconductivity\,\cite{jiang2022enhancing, zhang2022enhancement, peng2023enhanced, zhou2023robust}. Moreover, experimental data from underdoped and optimally doped cuprates suggest that long-range effective interactions, especially attractive ones, are necessary to fully explain the quasi-circular patterns observed in x-ray scattering\,\cite{boschini2021dynamic, scott2023low}. With these considerations in mind, we extend the Hubbard model to include EPCs and investigate their influence on the extremely overdoped CDW.

\subsection{Hubbard-Holstein Model with Local Coupling}\label{sec:phonon:HubbardHolstein}

Building on the Hubbard model findings depicted in Fig.~\ref{fig:Hubbard}, we incorporate local e-ph coupling by adopting the Hubbard-Holstein model, as formulated in Eq.~\eqref{eq:HubbardHolstein}. Our previous analysis of the Hubbard model identified a peak of charge susceptibilities at around 50\% doping. This leads us to focus on how phonon interactions alter the momentum distribution of these susceptibilities. Due to the computational complexity posed by the phonon degrees of freedom [see Appendix \ref{app:fermionSignAutoCorr}], we limit our simulations to a $6\times6$ square lattice in this section, corresponding to the smallest square cluster capable of accommodating the experimentally observed $(\pi/3,0)$ momentum. 

In Fig.~\ref{fig:HubbardHolstein}(a), we analyze the evolution of charge susceptibilities in the 50\% doped Hubbard-Holstein model across all momenta as the onsite EPC $g$ varies. Although typical EPC strengths in cuprate materials are around $0.5t$\,\cite{gvalue}, we extend our simulations up to $g=2t$ to test the generality of our conclusions. As expected, increasing $g$ amplifies the overall $\chi_c(\mathbf{q},\omega=0)$ at all momenta, driven by the direct coupling to the local electron density. Nevertheless, the peak intensity consistently appears at large momenta near $(\pi,\pi)$, with no significant enhancement at $(\pi/3,0)$, the momentum relevant to experimental observations.

We further explore the doping dependence of charge susceptibility distribution by fixing the coupling strength at $g=t$. As shown in Fig.~\ref{fig:HubbardHolstein}(b), even with such a strong EPC, the phase diagram remains qualitatively unchanged from that of the pure Hubbard model. The system continues to display a stripe state below 20\% doping, characterized by a peak at antinodal wavevector. Given the resolution limits of our $6\times6$ cluster, we can only identify the stripe wavevector at $(\pi/3,0)$, while it shifts with doping\,\cite{karakuzu2022stripe}. However, the primary focus of this study is in the extremely overdoped regime beyond 30\% doping. In this regime, checkerboard fluctuations at $(\pi,\pi)$ remains dominant, despite the presence of strong EPC. There is signs of the unidirectional CDW instability throughout the entire regime. Therefore, the onsite EPC, as considered in the Hubbard-Holstein model, still cannot account for the experimentally observed overdoped CDW at $(\pi/3,0)$. While the simulations are performed at a relatively high temperature, this conclusion is further supported by zero-temperature results discussed later. 

Although the inclusion of the onsite EPC does not resolve the issue of the $(\pi/3,0)$ CDW at the extremely overdoped regime, the doping dependence of charge susceptibilities  peaks near 50\% doping [see Fig.~\ref{fig:HubbardHolstein}(c)], consistent with the experimental observations\,\cite{li2023prevailing}. As discussed in Sec.~\ref{sec:Hubbard}, this monotonic doping dependence is driven by the strong Hubbard interaction, which is not altered by the inclusion of EPC.

\subsection{Charge Instability Driven by Nonlocal Coupling}

\begin{figure*}[!t]
\begin{center}
\includegraphics[width=\textwidth]{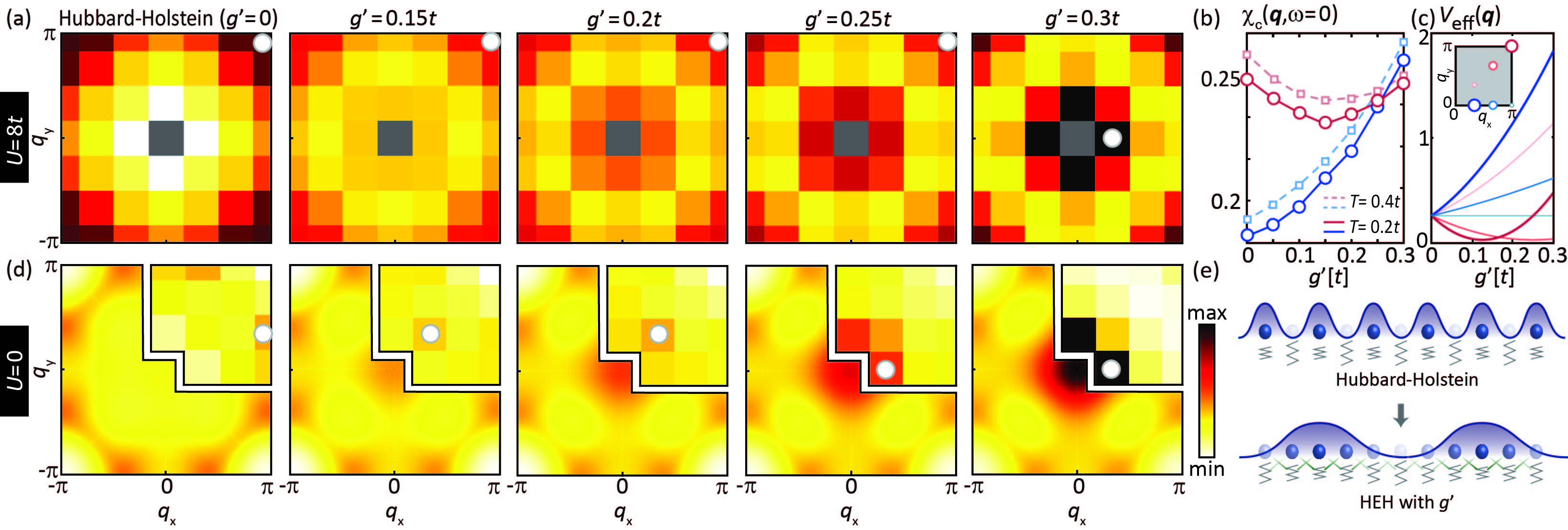}\vspace{-4mm}
\caption{\label{fig:HubbardExtendedHolstein}
(a) Charge susceptibility for the quarterly filled HEH model ($U=8t$ and $g=0.5t$) with different $g^{\prime}$s and $T=0.2t$. (b) The $\mathbf{q}=(\pi/3,0)$ (blue) and $(\pi,\pi)$ (red) susceptibilities for $T=0.4t$ (dashed curves) and $T=0.2t$ (solid curves) as a function of the $g^{\prime}$ with $U=8t$.The errorbar is smaller than the symbol. (c) Phonon-mediated effective interaction $|V_{\rm eff}(\mathbf{q})|$ for $\omega=0$ as a function of $g^\prime$ at specific momentum points marked in the inset. (d) The $\chi_c(\mathbf{q})$ obtained by RPA (lower-left) and DQMC (upper-right insets) simulation for the extended-Holstein model with $U=0$ and $T=0.2t$. (e) Schematic illustrating that the nonlocal EPC $g^\prime$ favors longer period charge modulation, while onsite EPC $g$ favors short period charge modulation.\vspace{-1mm}
}
\end{center}
\end{figure*}

Our analysis of the Hubbard-Holstein model reveals that the inclusion of onsite EPC within realistic coupling strengths does not substantially alter the charge susceptibility patterns observed in the pure Hubbard model. To investigate the origin of the experimentally observed overdoped CDW, we extend our model to nonlocal EPCs\,\cite{hebert2019one}, which are known to stabilize polarons under weak coupling conditions\,\cite{perroni2004polaron, fehske2000lattice, alexandrov1999mobile}. These couplings have been found essential in explaining the attractive interactions recently observed in 1D cuprates\,\cite{chen2021anomalously, wang2021phonon, tang2023traces}. Considering that electrostatic couplings, like those to apical oxygens, decay with distance\,\cite{wang2021phonon}, our primary focus is on the EPC between nearest-neighbor electron density and lattice displacement, labeled as $g^\prime$ in Eq.~\eqref{eq:HubbardExtendedHolstein}. Starting with the quarter-filled Hubbard-Holstein model discussed in Sec.~\ref{sec:phonon:HubbardHolstein} ($g^\prime=0$), we gradually increase $g^\prime$ while holding $g=0.5t$ [see Fig.~\ref{fig:HubbardExtendedHolstein}(a)]. The introduction of nearest-neighbor EPC results in the emergence of small-momentum charge susceptibility at the zone center. At the same time, the checkerboard charge fluctuations at large momenta persist. Once $g^\prime$ exceeds $\sim0.25t$, the small-wavevector $\chi_{\rm c}$ at $\mathbf{q}=(\pi/3,0)$ surpasses the $(\pi,\pi)$ susceptibility in intensity, becoming the dominant wavevector [see Fig.~\ref{fig:HubbardExtendedHolstein}(b)]. This critical coupling strength is slightly above the estimated value from the octahedral symmetry, where $g^\prime\sim g/\sqrt{5}$\,\cite{wang2021phonon}, but the Jahn-Teller effect may bring $g^\prime$ and $g$ closer in cuprate materials, making the simulated scenario realistic. 

Importantly, the intensities at these key wavevectors and their doping dependence are resilient to temperature changes [see Fig.~\ref{fig:HubbardExtendedHolstein}(b)], consistent with RXS results\,\cite{li2023prevailing}. (See Sec.~\ref{sec:phonon:temperature} for a more detailed demonstration of the resilience against temperature change.) Furthermore, while these nonlocal EPCs shift the dominant wavevector, they do not alter the doping dependence of the charge susceptibility intensities, which remains maximal near quarter filling (see the Appendix \ref{app:HEH:doping}).

The emergence of small-momentum susceptibilities can be elucidated by analyzing the phonon-mediated effective interactions between electrons\,\cite{varelogiannis1996density}. The nearest-neighbor coupling $g^\prime$ modulates the EPC in momentum space, i.e.,~$g_\mathbf{q}=g+2g^\prime(\cos q_x+\cos q_y)$. When $g^\prime$ shares the same sign as $g$, the coupling is predominantly projected onto small momenta. As the phonon-mediated dynamical interaction $V_{\rm eff}(\mathbf{q},\omega)={g_\mathbf{q}^2}/{M(\omega^2-\omega_{\rm ph}^2)}$ scales with $g_\mathbf{q}^2$, this modulation further affects the charge susceptibility near the zone center, including the anticipated $(\pi/3,0)$ [see Figs.~\ref{fig:HubbardExtendedHolstein}(b) and (c)]. Simultaneously, the momentum dependence of the attractive $V_{\rm eff}$ also adjusts the $(\pi,\pi)$ charge instability. As $g_{\mathbf{q}=(\pi,\pi)}$ decreases and eventually turns negative with increasing $g^\prime$, the $V_{\rm eff}\propto g_\mathbf{q}^2$ exhibits a nonmonotonic trend with respect to $g^\prime$. Such a nonmonotonic dependence is also reflected in the evolution of $\chi_{\rm c}(\mathbf{q}=(\pi,\pi))$. Alternatively, the impact of nonlocal EPC can be interpreted in the real-space picture. As shown in Fig.~\ref{fig:HubbardExtendedHolstein}(e), the presence of $g^\prime$ clusters electrons around individual lattice displacements, transforming short-range doublon-hole fluctuations into form a longer-range charge pattern.

The relationship between $V_{\rm eff}$ and small-momentum charge instabilities suggests an independent origin for these intensities, allowing them to coexist with large-momentum doublon-hole fluctuations induced by $U$. These two interactions manifest minimal overlap in momentum space. As a demonstration, we set Hubbard $U$ to zero in the insets of Fig.~\ref{fig:HubbardExtendedHolstein}(d) and find that the DQMC-simulated charge susceptibility intensifies only near the zone center, with negligible intensity at $(\pi,\pi)$ compared to that of the HEH model [i.e., Fig.~\ref{fig:HubbardExtendedHolstein}(a)]. These small-momentum susceptibilities are consistent in models with and without $U$ when the EPC strength is the same. The Hubbard $U$ plays the role of determining the overall charge susceptibility and its doping dependence, necessary for achieving a maximum at 50\% doping.

While the dominant wavevector for charge susceptibility has reached the experimentally observed $(\pi/3,0)$ in Fig.~\ref{fig:HubbardExtendedHolstein}(a), we caution against over-interpreting the quantitative value of this wavevector, given that it represents the smallest nonzero wavevector in the $6\times 6$ square lattice. As discussed in the Appendix \ref{app:finiteSize}, simulations of larger systems, albeit at higher temperatures, suggest a dominant wavevector shifting closer to $(0,0)$. Moreover, the phonon-mediated $|V_{\rm eff}|$ is unlikely to peak at nonzero wavevectors if only positive onsite and nearest-neighbor couplings ($g$ and $g^\prime$) are considered. This limitation necessitates the exploration of a parameter regime incorporating longer-range couplings, such as $g^{\prime\prime}$, $g^{\prime\prime\prime}$, and  $g^{\prime\prime\prime\prime}$) in Eq.~\eqref{eq:HubbardExtendedHolstein}. These extended couplings, which originate from the electrostatic nature of the site-phonon interactions, significantly modulate the momentum-space EPC $g_\mathbf{q}$, thus shaping the small-momentum charge distributions more precisely.

\begin{figure}[!b]
\begin{center}
\includegraphics[width=8.5cm]{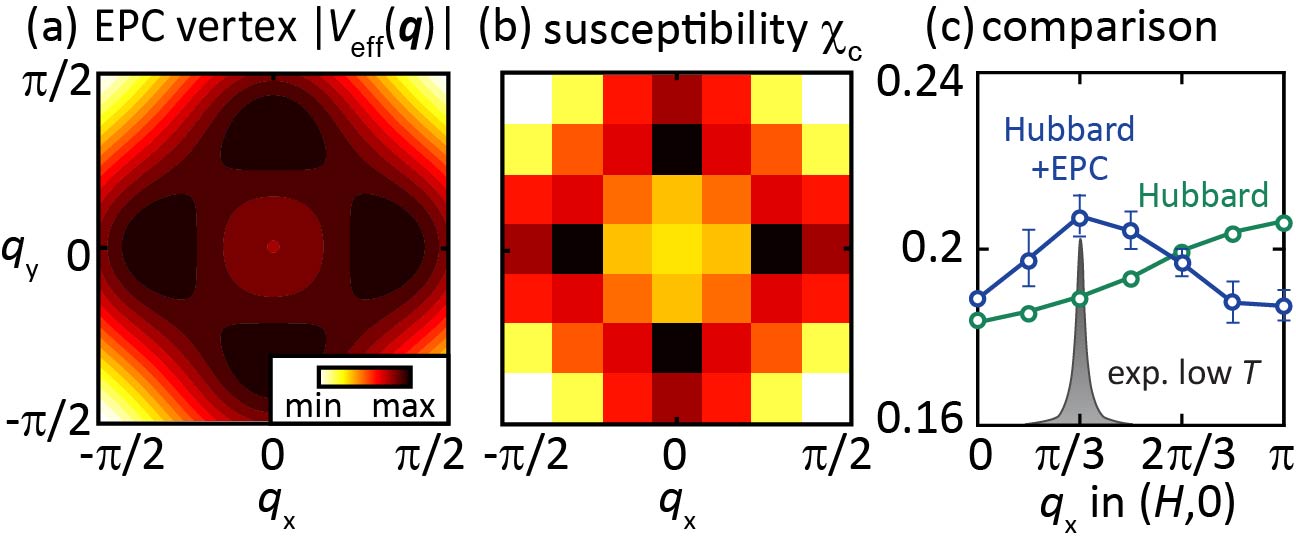}\vspace{-4mm}
\caption{\label{fig:effectiveV}
(a) Effective dynamical interaction $|V_{\rm eff}(\mathbf{q},\omega=0)|$ mediated by phonons with longer-range couplings ($g^{\prime}=0.55g$, $g^{\prime\prime}=-0.1g$, $g^{\prime\prime\prime}=g^{\prime\prime}/\sqrt{2}$, $g^{\prime\prime\prime\prime}=g^{\prime\prime}/2$). (b) DQMC-simulated charge susceptibility $\chi_{\rm c}$ for quarterly filled HEM ($U=8t$ and $g=0.5t$) obtained at $T=0.25t$. (c) Comparison of the background-removed RXS results at 31\,K obtained adapted from Ref.~[\onlinecite{li2023prevailing}] (gray peak), the simulated charge susceptibility for the Hubbard model (green), and that for the HEH model (blue) along the antinodal direction.\vspace{-4mm}}
\end{center}
\end{figure}

To quickly estimate phonon-induced CDW orders without system-size limitation, we utilize the random phase approximation (RPA) to map the momentum distribution of $\chi_{\rm c}(\mathbf{q})$, employing $V_{\rm eff}(\mathbf{q},\omega)$ as the interaction vertex. This method proves effective, as shown by comparing the RPA (major parts) and DQMC (insets) in Fig.~\ref{fig:HubbardExtendedHolstein}(c). The RPA successfully captures small-momentum instabilities in systems with $U=8t$, as the EPC and electronic interactions determine small- and large-momentum susceptibilities almost independently. 

Due to the screening effects at long distances, we slightly relax the assumption that EPC decays strictly as $1/r$ and instead adopt the parameters in Fig.~\ref{fig:effectiveV}, where $|V_{\rm eff}(\mathbf{q},\omega=0)|$ peaks at $(\pi/3,0)$. Using this parameter set, we simulate $\chi_{\rm c}$ using DQMC. As expected, the charge wavevector is precisely pinned at $(\pi/3,0)$ for $12\times 12$ systems, validating the potential of inducing a robust, experimentally consistent charge order through long-range EPC. While these couplings are tied to materials' crystal and electronic structure, slight deviations from this commensurate wavevector may be mitigated by domain boundaries and disorders in real materials\,\cite{campi2015inhomogeneity}. Our simulation ignores the long-range electronic Coulomb repulsion, which typically contributes an additional $\sim 1/|\mathbf{q}|^2$ interaction as its bare form. However, the single-band Hubbard model already incorporates corrections from these Coulomb interactions during the projection from its multi-band prototype\,\cite{hybertsen1990renormalization,hansmann2014importance}. As a result, the effective single-band wavefunction combines the copper and oxygen components\,\cite{zhang1988effective}, where the in-plane Coulomb interaction is largely screened by the copper-oxygen covalent bond in the extremely overdoped regime. Such a screening is more severe in the overdoped regime with a large Fermi surface. 

\subsection{Temperature Dependence}\label{sec:phonon:temperature}
\begin{figure}[!b]
    \centering
    \includegraphics[width=\columnwidth]{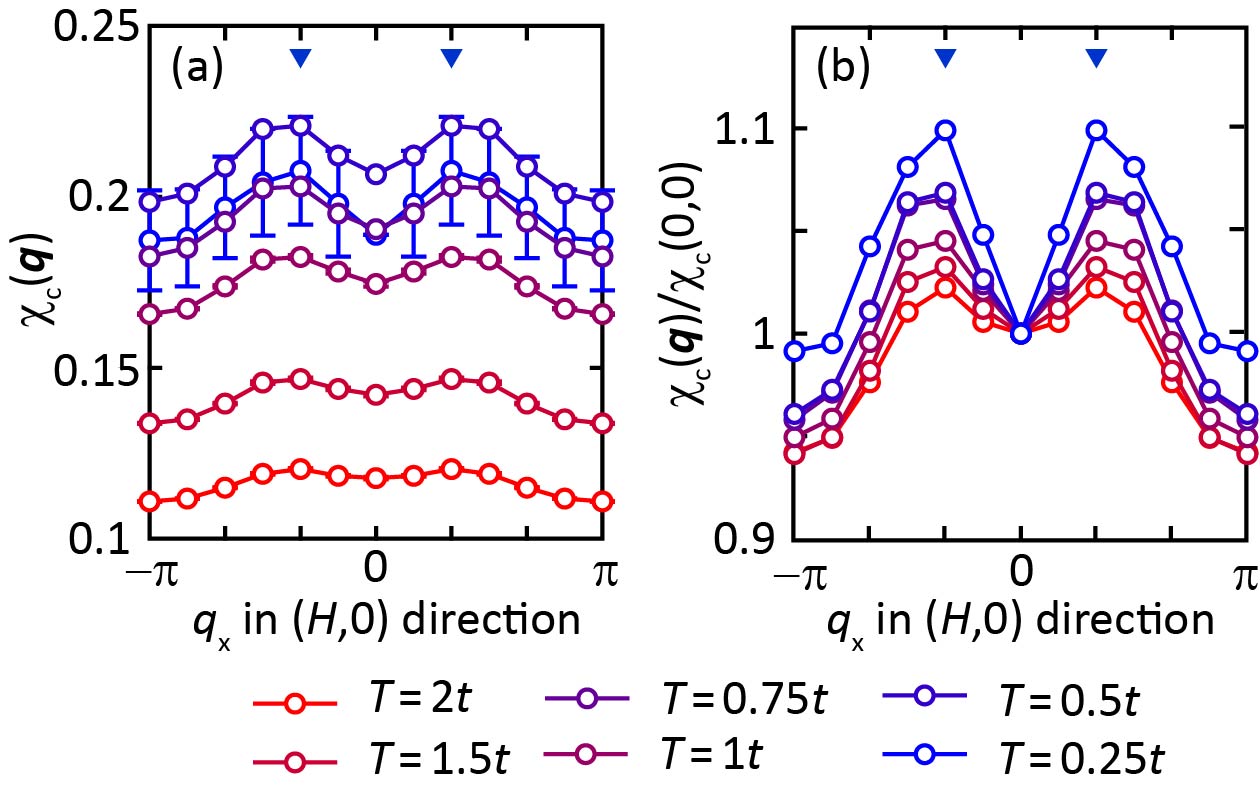}\vspace{-5mm}
    \caption{\label{fig:temperatureDep}
    (a) Charge susceptibility and (b) normalized susceptibility relative to $\chi(0,0)$ along the $(H,0)$ direction for different temperatures. All model parameters are identical to those in Fig.~\ref{fig:effectiveV}. The triangles mark the wavevector observed in experiment\,\cite{li2023prevailing}. Due to significant particle number fluctuations, the $\mathbf{q}=(0,0)$ charge susceptibility at $T=0.25t$ is extrapolated. Error bars in (b) are not shown.
}    
\end{figure}
The fermion sign problem in DQMC limits our ability to simulate at low temperatures comparable to experimental conditions (below 300\,K). To extrapolate the potential impact of temperature, we analyze how susceptibilities change at higher temperatures. Fig.~\ref{fig:temperatureDep}(a) presents the results using the same model parameters as in Fig.~\ref{fig:effectiveV}, but at elevated temperatures. The $(\pi/3,0)$ peak in $\chi_c$ is consistently observed throughout all temperatures studied. This behavior becomes more apparent when susceptibilities at different temperatures are normalized to the intensity at the zone center $\mathbf{q}=(0,0)$, as shown in Fig.~\ref{fig:temperatureDep}(b). While the relative peak height diminishes as temperature increases, the peak's position remains unchanged. Although this observation does not definitively prove that this charge order persists at low temperatures beyond the DQMC's reach, it does provide strong indirect evidence that the charge order is resilient to temperature changes.

\begin{figure}[!t]
    \centering
    \includegraphics[width=\columnwidth]{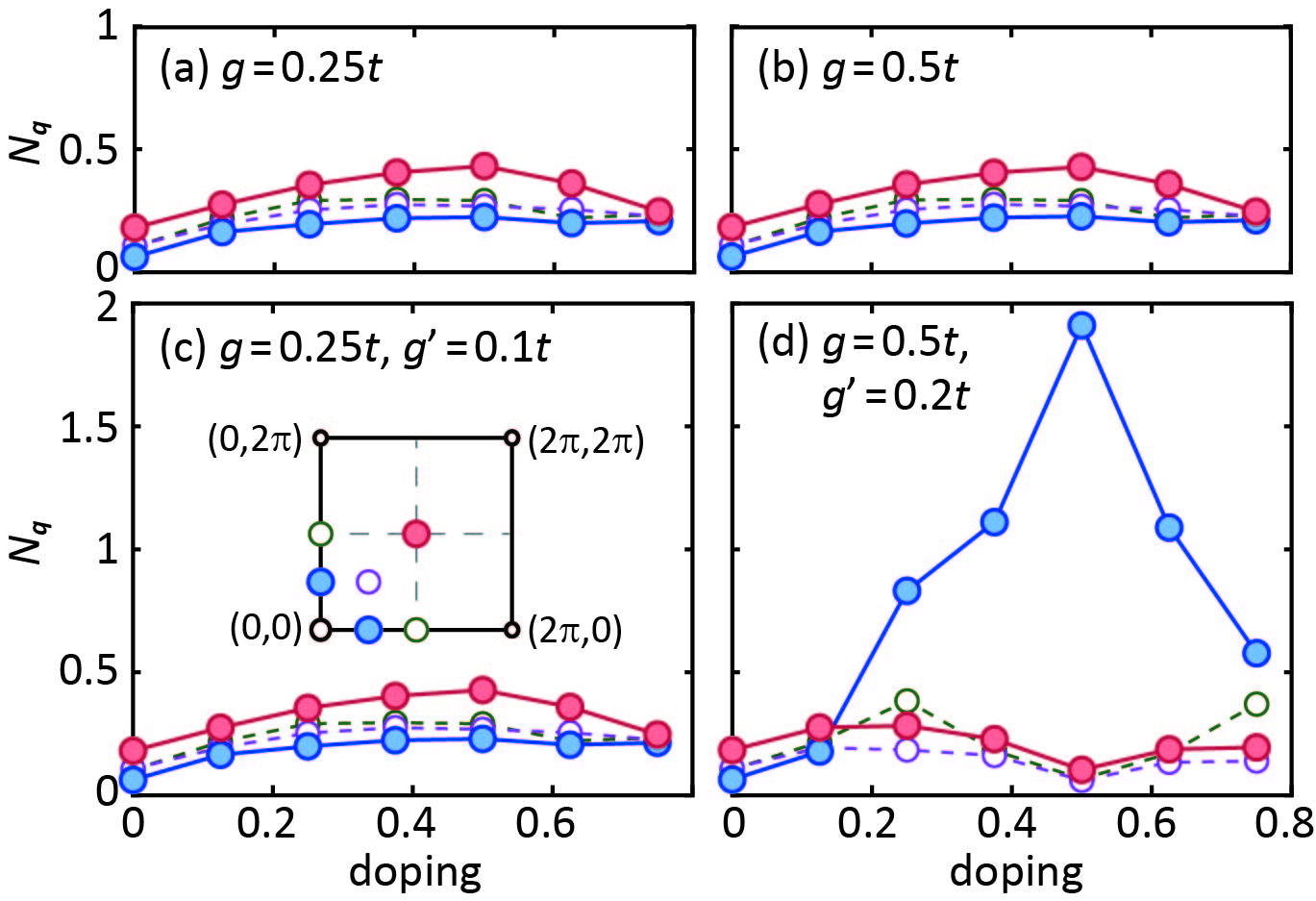}\vspace{-3mm}
\caption{\label{fig:NGSED}
    Static charge structure factor $N_\textbf{q}$ in the HEH model for momenta accessible in a $4\times4$ cluster, simulated using NGSED. The model parameters used are $U=8t$, $t^\prime=-0.15t$, and (a) $g=0.25t$, (b) $g=0.5t$, (c) $g=0.25t$, $g^\prime =0.1t$, and (d) $g=0.5t$, $g^\prime =0.2t$.
}   
\end{figure}

To further validate our results at extremely low temperatures comparable to experimental conditions, we leverage the non-Gaussian exact diagonalization (NGSED) method, a zero-temperature technique, to simulate the HEH models. NGSED is a variational approach built upon exact diagonalization (ED), designed to address systems with both strong electronic correlations and EPCs. Generally, any electron-phonon wavefunction can be decomposed as 
\begin{eqnarray}\label{eq:wvfuncansatzNGS}
    \big|\Psi\big\rangle  = U_{\rm NGS}(\{\lambda_\mathbf{q}\}) |\psi_{\rm ph}\rangle \otimes|\psi_{\rm e}\rangle\,.
\end{eqnarray}
Here, the electron-phonon entanglement is encapsulated by the non-Gaussian transformation $U_{\rm NGS}=e^{i\mathcal{S}[c^\dagger, c, a^\dagger, a]}$, where $\mathcal{S}$ is a polynomial involving at least two electronic creation and annihilation operators and at least one phonon operator. While this method approaches exactness as the order in $\mathcal{S}$ increases, previous numerical benchmarks using ED and DQMC on small clusters have shown that truncating to the lowest-order terms is sufficient for site-phonons\,\cite{wang2020zero, wang2021fluctuating}. Thus, the non-Gaussian transformation  reduces to a generalized polaronic transformation:
\begin{equation}\label{eq:NGStransform}
    U_{\rm NGS}(\{\lambda_\mathbf{q}\}) \approx e^{i-\frac{1}{\sqrt{N}}\sum_{\mathbf{q} i } \lambda_\mathbf{q}  e^{i\mathbf{q}\cdot \mathbf{r}_i} (a_\mathbf{q}-a_{-\mathbf{q}}^\dagger) n_{i}}\,,
\end{equation}
where $\rho_{\mathbf{q}}=\sum_{i} n_{i}e^{-i \mathbf{q}\cdot \mathbf{r}_i}$ denotes the momentum-space electron density, and $p_{\mathbf{q}}=\sum_{i} P_{i}  e^{-i \mathbf{q}\cdot \mathbf{r}_i} /\sqrt{N}$ represents the phonon momentum operator. Within the wavefunction ansatz described in Eq.~\eqref{eq:wvfuncansatzNGS}, the ground state solution is determined by minimizing the total energy in the variational parameter space spanned by $\{\lambda_\mathbf{q}\}$, $|\psi_{\rm ph}\rangle$, and $|\psi_{\rm e}\rangle$. The NGSED method self-consistently optimizes the variational parameters and the individual states, converging to the ground state\,\cite{wang2020zero}. Due to the complexity of evaluating excited states within this self-consistent framework, we restrict our focus to the static (equal-time) charge structure factor $N_\mathbf{q}$ in the NGSED simulation, using it as a metric for the charge instability.

As shown in Fig.~\ref{fig:NGSED}, in the Hubbard-Holstein model with no or minimal $g^\prime$, $(\pi,\pi)$ dominates over other momenta at zero temperature. However, as $g^\prime$ is increased to $0.2t$, there is a notable enhancement in charge correlation at $(\pi/2,0)$, with this enhancement peaking at 50\% doping. It should be noted that due to the constraints of a finite-size cluster (a $4\times4$ square lattice here), the NGSED simulation is unable to reach the expected $(\pi/3,0)$ momentum, leaving $(\pi/2,0)$ as the closest momentum point accessible. Nevertheless, the distinct behavior with and without $g^\prime$ are consistent with our DQMC results discussed above, suggesting that the conclusions drawn the EPC vertex are robust even at zero temperature.

\subsection{Impact of Phonon Frequencies and Band Structure}
The constraints of DQMC simulations, particularly the severe auto-correlation issues caused by low-frequency phonons, necessitated our choice of a phonon frequency $\omega_{\rm ph}=t$ for the results discussed above. However, realistic phonon frequencies in cuprates can be as low as $60-70$\,meV\,\cite{vishik2014angle,he2018rapid}, suggesting that $\omega_{\rm ph}\sim 0.2t$ is more accurate for LSCO. Despite adopting a higher phonon frequency, our findings suggest that the wavevector of the extremely overdoped charge order near the Brillouin zone center is primarily determined by the momentum distribution of the effective EPC vertex $g_{\mathbf{q}}^2/\omega_{\rm ph}^2$. This conclusion is supported by both our DQMC simulations and comparisons with RPA in Sec.~\ref{sec:phonon:HubbardHolstein}. Therefore, when phonons are optical with minimal frequency dependence (i.e.,~$\omega_\mathbf{q}\approx \omega_{\rm ph}$), the influence of phonon frequency on the relative distribution of charge susceptibility in momentum space is largely quantitative rather than qualitative.

\begin{figure}[!t]
    \centering
    \includegraphics[width=\columnwidth]{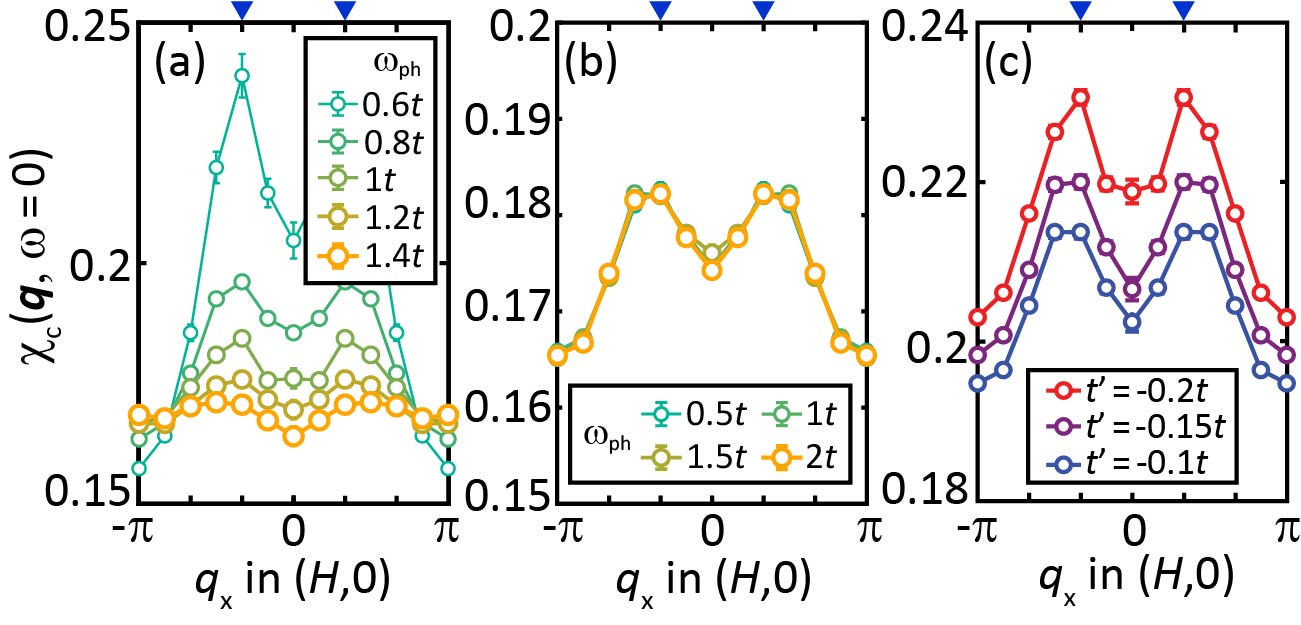}\vspace{-4mm}
\caption{\label{fig:omegatprimeDep}
    (a,b) Dependence of charge susceptibilities on phonon frequency by fixing (a) all EPC strengths $g$, $g^\prime$, etc and (b) the ratio $g/\omega_{\rm ph}$. (c) Dependence of the charge susceptibility on the band structure in the quarter-filled HEH model. All other model parameters are identical to those in Fig.~\ref{fig:effectiveV} and the temperature is $T=0.5t$. The triangles mark the wavevector observed in experiment\,\cite{li2023prevailing}.
}
\end{figure}

To demonstrate this, we simulate $\chi(\mathbf{q},\omega=0)$ for various phonon frequencies at a high temperature of $T=1t$. When the EPC $g_\mathbf{q}$ is kept constant, decreasing $\omega_{\rm ph}$ leads to a uniform increase in the $(\pi/3,0)$ susceptibility, driven by the enhanced vertex [see Fig.~\ref{fig:omegatprimeDep}(a)]. This simple scaling behavior of the phonon frequency is further demonstrated by fixing the ratio $g_\mathbf{q}/\omega_{\rm ph}$ while varying $\omega_{\rm ph}$, thereby preserving the vertex amplitude $g_\mathbf{q}^2/\omega_{\rm ph}^2$. As shown in Fig.~\ref{fig:omegatprimeDep}(b), the susceptibility distribution remains largely unaffected across different frequencies. Therefore, employing a single phonon frequency $\omega_{\rm ph}=t$ in most simulations is sufficient to identify the dominant wavevector and relative susceptibility distribution.

In addition to the phonon frequency, the electronic band structure is a crucial factor that can significantly influence collective excitations by modifying the electronic Fermi surface shape, a characteristic that varies across different cuprate materials. As discussed in Sec.~\ref{sec:model}, this band structure is parameterized by the next-nearest-neighbor hopping $t^\prime$. To test the robustness of our conclusions across a wider range of cuprate materials, we slightly adjust $t^\prime$ from the accepted LSCO value of $t^\prime=-0.15t$, while keeping all other parameters the same as in Fig.~\ref{fig:effectiveV}. As shown in Fig.~\ref{fig:omegatprimeDep}(c), this variation of $t^\prime$ within $\pm 0.05t$ leads to an approximate 5\% change in the overall charge susceptibilities. However, despite this quantitative scaling, the dominant wavevector consistently remains at $(\pi/3,0)$, reflecting the robustness of this extremely overdoped charge order against the changes in specific band parameters when nonlocal EPCs are present. These results further support the generality of our conclusions regarding the EPC-vertex-determined charge order across the various cuprate materials, consistent with the observed experimental signatures observed in both LSCO and Bi2201\,\cite{peng2018re, li2023prevailing}.

\section{Potential Contributions from Bond Phonons}\label{sec:nonHolstein}
In the previous sections, we concentrated on site-phonons that couple to electronic density, as described by the Hamiltonians expressed in Eqs.~\eqref{eq:HubbardHolstein} and \eqref{eq:HubbardExtendedHolstein}. However, in cuprate materials, electrons also couple to bond phonons, where the coupling matrix elements are dependent on the electronic momentum. To extend our analysis, we investigate various bond phonons and their impacts on charge instabilities in the extremely overdoped regime. The electron-phonon coupled part of Hamiltonian for a general Fr\"{o}hlich EPC system is
\begin{equation}
    \mathcal{H}_{\rm e-ph}=\sum_{{\bf k},{\bf q},\sigma}\frac{g_{{\bf k}{\bf q}}}{\sqrt{N}} c_{{\bf k}-{\bf q}\sigma}^\dagger c_{{\bf k},\sigma}(a_{\bf q}^\dagger+a_{-{\bf q}})+\omega_{\rm ph}\sum_{\bf q} a_{\bf q}^\dagger a_{\bf q}\,.
\end{equation}
Building upon the conclusions from Sec.~\ref{sec:phonon}, we focus on their influence on the EPC vertex within the RPA framework, under the assumption that large-momentum checkerboard fluctuations induced by electronic correlations do not affect the zone-center charge susceptibility in the extremely overdoped regime. In this framework, the charge susceptibility is given by
\begin{eqnarray}
\chi({\bf p},i\omega_p;{\bf q},i\omega_q)=\frac{\chi_0({\bf p},i\omega_p;{\bf q},i\omega_q)}{1-P({\bf p},i\omega_p;{\bf q},i\omega_q)}\,,
\end{eqnarray}
where $\chi_0$ denotes the bare susceptibility of non-interacting electrons, and 
\begin{eqnarray}
P(\mkern-2mu{\bf p},\mkern-2mui\omega_p;{\bf q},\mkern-2mui\omega_q\mkern-2mu)\mkern-2mu=\mkern-2mu\sum_{{\bf k}} \mkern-2mu\frac{g_{{\bf k q}}g_{{\bf p},{\bf -q}}\omega_{\rm ph}}{N(\omega_q^2\mkern-2mu+\mkern-2mu\omega_{\rm ph}^2)}  \frac{n\mkern-2mu_F(\mkern-1mu\epsilon_{\bf p}\mkern-1mu)\mkern-2mu-\mkern-2mun\mkern-2mu_F(\mkern-1mu\epsilon_{{\bf p}\mkern-2mu+\mkern-2mu{\bf q}}\mkern-1mu)}{i\omega_q+\epsilon_{\bf p}-\epsilon_{{\bf p}+{\bf q}}}\,. 
\end{eqnarray}
Here, spin indices are omitted, and $n_F$ represents the Fermi-Dirac distribution. Finally, the charge susceptibility is calculated as $\chi({\bf q},i\omega_q)=\sum_{{\bf p},\omega_p}\chi({\bf p},i\omega_p;{\bf q},i\omega_q)/\beta N$. Consistent with previous discussions, we focus on the zero-frequency charge susceptibility $\chi({\bf q},\omega=0)$, which acts as a fingerprint for charge order.

\begin{figure}[!t]
\begin{center}
\includegraphics[width=\columnwidth]{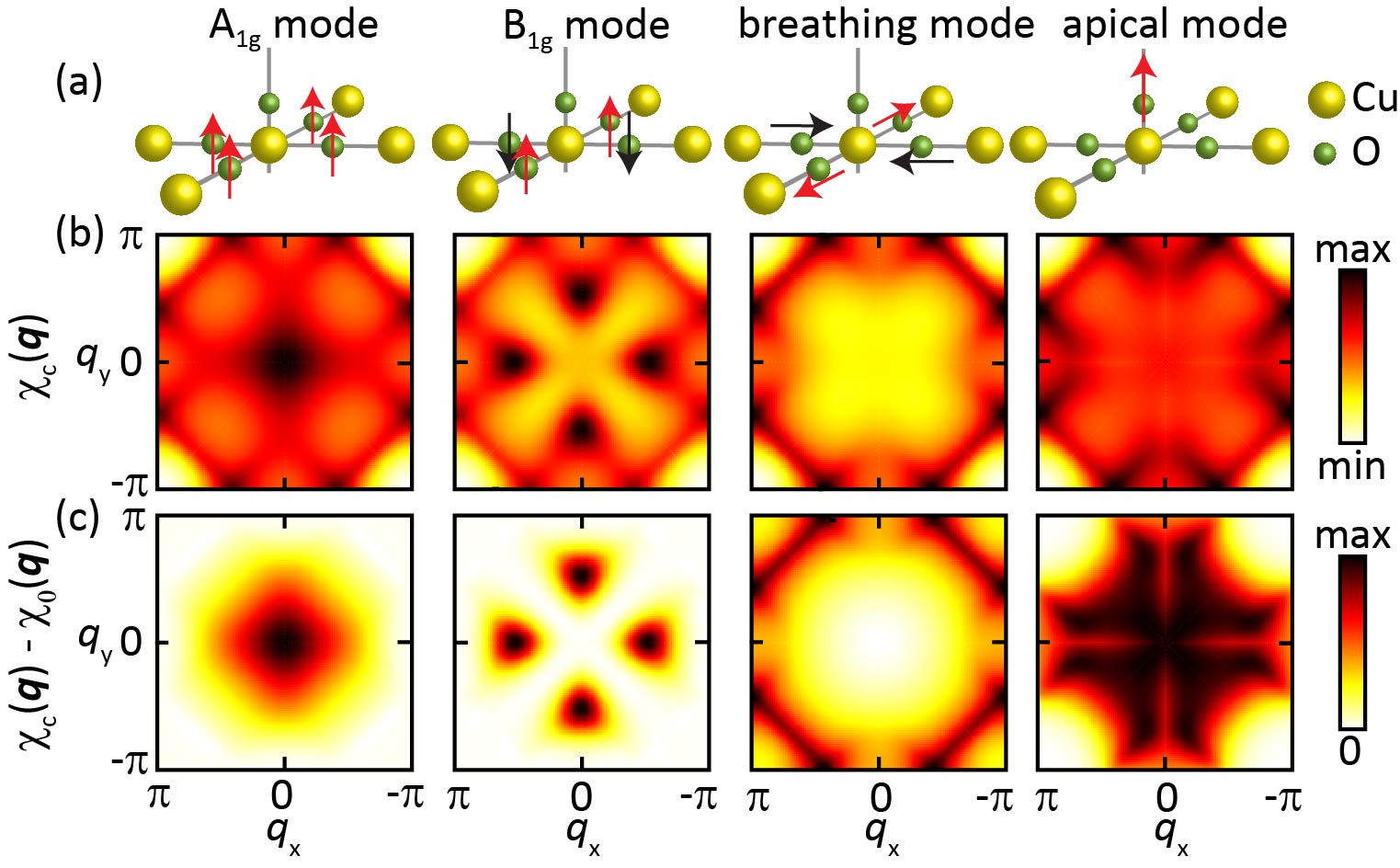}\vspace{-3mm}
\caption{\label{fig:bondPhonon}
    (a) Schematics illustrating bond-phonon modes in the cuprates. (b) The RPA charge susceptibility $\chi_{\rm RPA}(\mathbf{q})$ for each phonon mode. (c) The difference in charge susceptibility from the Lindhard response, revealing phonon-induced instabilities. Simulations are performed on a 50\% doped $128\times 128$ square lattice at $T=0.1t$.
}
\end{center}
\end{figure}

We examine several significant types of bond phonons in cuprates by projecting their couplings onto the Zhang-Rice singlet electronic wavefunctions\,\cite{johnston2010systematic}. As shown in Fig.~\ref{fig:bondPhonon}, the breathing phonon predominantly enhances charge susceptibility near the Fermi momentum, while the $\rm A_{\rm 1g}$ phonon primarily induces charge instability around the $\Gamma$ point. The apical phonon, on the other hand, generates a broad range of fluctuations along the nodal direction. Interestingly, the $\rm B_{1g}$ phonon leads to susceptibilities in the antinodal direction and may contribute to the $(\pi/3,0)$ charge order observed in extremely overdoped LSCO, albeit with a slightly different wavevector. The $\rm B_{1g}$ phonon has been shown to support $d$-wave superconductivity in cuprates\,\cite{johnston2010systematic,he2018rapid,honerkamp2007phonons}. The impact of these bond phonons on superconductivity, particularly in the extremely overdoped regime, is beyond the scope of this paper but warrants further study, particularly in understanding the role of the extremely overdoped CDW in high-$T_c$ systems.

\section{Summary and Outlook}\label{sec:conclusion}
In conclusion, we have conducted a thorough investigation of the charge instabilities in the extremely overdoped cuprates, utilizing the Hubbard model and its extensions with phonons.  Excluding Fermi-surface instability, the extremely overdoped Hubbard model successfully captures the nonmonotonic doping dependence of the experimentally identified period-6 CDW, reflecting the persistence of remnant correlations even in the overdoped regime. However, the limitations of the Hubbard model and the  Hubbard-Holstein model with local EPCs become evident when attempting to address the $(\pi/3,0)$ wavevector, as local particle-hole fluctuations at $(\pi,\pi)$ dominate. By introducing nonlocal EPCs, with coupling strengths determined by geometric relations, we have provided an explanation for the period-6 CDW in extremely overdoped cuprates. Crucially, this phonon-induced CDW is shown to be robust against variations in temperature and specific model parameters. The phonon-mediated effective interaction pins the CDW at $(\pi/3,0)$ across a wide range of doping levels, consistent with RXS experiments. 

Since the extremely overdoped regime is well-separated from the pseudogap and AFM phases, where spin fluctuations dominate, the phonon-driven nature of the CDW suggests a subleading interaction in cuprates. This interaction is overwhelmed by the Hubbard $U$ in underdoped and optimally doped cuprates, as detailed in this paper. However, recent studies increasingly indicate that phonons, alongside strong correlations, contribute significantly to $d$-wave superconductivity\,\cite{he2018rapid,jiang2022enhancing, peng2023enhanced, wang2022robust,cai2023high}. In the overdoped regime, where the impact of correlations is separable, the EPC identified here is crucial for the theory of understanding unconventional superconductivity, particularly given its BCS-like nature at the emergence from the overdoped side\,\cite{he2018rapid}. 

\section{Acknowledgement}
We thank the experimental inputs from Yingying Peng and insightful discussions from Ilya Esterlis. This work is supported by the Air Force Office of Scientific Research Young Investigator Program under grant FA9550-23-1-0153. This research used resources of the National Energy Research Scientific Computing Center, a DOE Office of Science User Facility supported by the Office of Science of the U.S. Department of Energy under Contract No.~DE-AC02-05CH11231 using NERSC award BES-ERCAP0027096.

\appendix

\section{Details about the Doping Dependence in the Hubbard Model}\label{app:Hubbard}
\begin{figure}[!b]
\centering
\includegraphics[width=\linewidth]{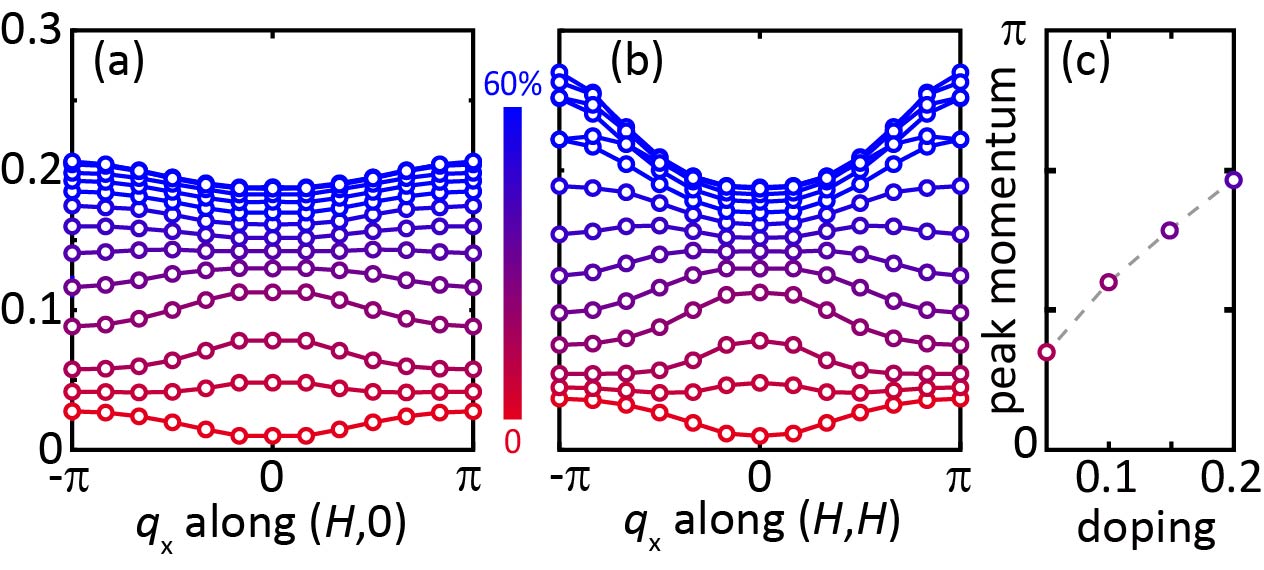}\vspace{-3mm}
\caption{\label{fig:appendix:HubbardCuts}
    (a,b) Charge susceptibilities along the (a) $(H,0)$ and (b) $(H,H)$ directions for the Hubbard model with $t^\prime=-0.15t$, $U=8t$, across different doping levels at $T=0.4t$. (c) Dominant wavevectors along the $(H,0)$ direction within the stripe regime, fitted through the multi-Lorentzian functions.
}
\end{figure}

In addition to the false-color representation in Fig.~\ref{fig:Hubbard} of the main text, we present here the antinodal and nodal cuts of the charge susceptibility for hole-doped Hubbard models and analyze the detailed evolution of these features with doping. As shown in Figs.~\ref{fig:appendix:HubbardCuts}(a) and (b), the system exhibits AFM at half-filling, with the subleading charge instability of doublon-hole fluctuations centered at $(\pi,\pi)$\,\cite{nowadnick2012competition}. As doping increases, stripe order emerges, shifting the dominant wavevector closer to the zone center. The fluctuating nature of the stripes causes significant momentum broadening, resulting in a greater spread of the charge susceptibility peaks compared to their separation. To identify the precise wavevector, we fit the susceptibility data using $C_4$-symmetric multi-Lorentzian functions, following Ref.~[\onlinecite{huang2023fluctuating}]. The fitted wavevectors are shown in Fig.~\ref{fig:appendix:HubbardCuts}(c). This fitting procedure works well for identifying stripes in the 5\%-20\% hole doping range, indicating the presence of stripes in the Hubbard model, but becomes less effective beyond this regime. 

\begin{figure}[!t]
    \centering
    \includegraphics[width=\linewidth]{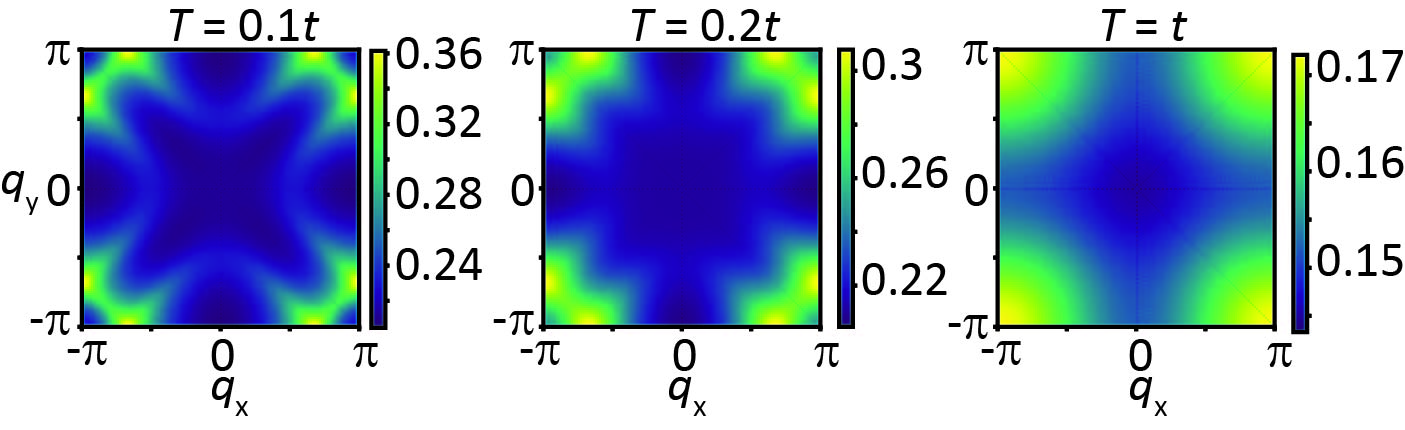}\vspace{-3mm}
    \caption{\label{fig:linhard}
Charge susceptibilities calculated from the Lindhard response functions for non-interacting electrons at 30\% hole doping and temperatures $T=0.1t$, $0.2t$ and $t$, respectively. The band structure parameters are identical to those used in Fig.~\ref{fig:Hubbard}, \textit{i.e.}, $t'=-0.15t$.
}
    
\end{figure}

Beyond 25\% doping, the state deviates significantly from the stripe states that smoothly evolve with increasing doping. This divergence is primarily indicated by a rapid increase in susceptibility near $(\pi,\pi)$, which overwhelms the instabilities in the antinodal direction and reflects a shift towards checkerboard fluctuations. In the extremely overdoped regime, these checkerboard fluctuations dominate over the entire Brillouin zone. This phenomenon can be attributed to two main reasons. Firstly, the Fermi surface geometry in the extremely overdoped regime naturally favors charge instability at large momenta near $(\pi,\pi)$, as shown by the Lindhard response function for non-interacting electrons in Fig.~\ref{fig:linhard}. Secondly, residual correlations within the system lead to doublon-hole fluctuations that further reinforce the $(\pi,\pi)$ charge fluctuations. The combination of these influences results in the observed dominance of $(\pi,\pi)$ excitations, while the $(\pi/3,0)$ charge order -- observed experimentally in Ref.~[\onlinecite{li2023prevailing}] -- is notably absent in the extremely overdoped Hubbard model.

\begin{figure}[!t]
\begin{center}
\includegraphics[width=\linewidth]{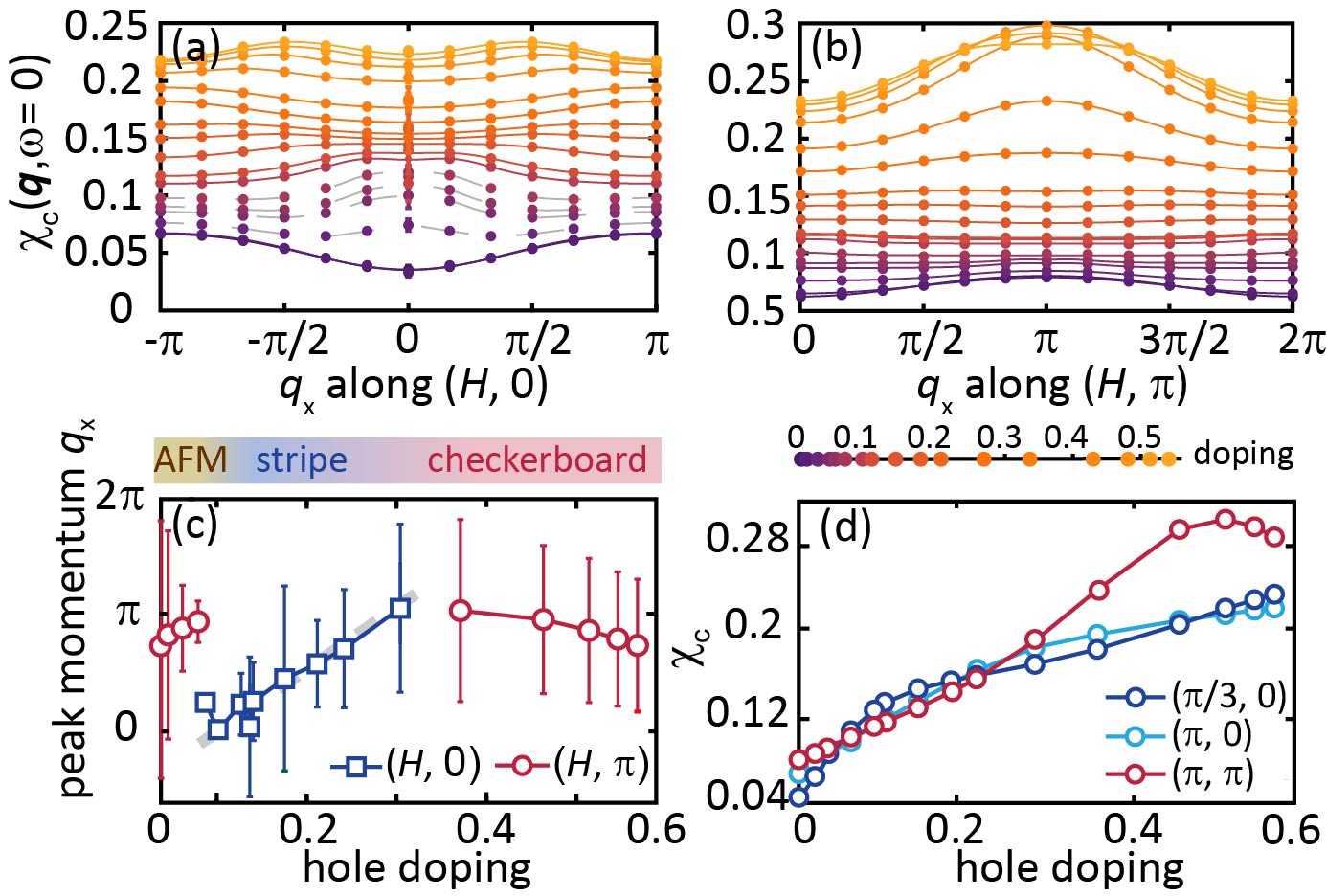}\vspace{-2mm}
\caption{\label{fig:HubbardRectangular}
    (a,b) Doping dependence of charge susceptibilities, along the (a) $(H,0)$ and (b) $(H,\pi)$ direction, obtained from the Hubbard model on a $12\times4$ rectangular lattice at $T=0.25t$. The model parameters are chosen as $U=6t$ and $t^\prime=-0.25t$, following Ref.~\,[\onlinecite{huang2023fluctuating}]. (c) Doping dependence of the CDW wavevectors along the $(H,0)$ (blue) or $(H,\pi)$ (red) directions, derived from double-Lorentzian fitting. The grey dashed curve highlights the linear doping dependence in the stripe phase, with all the three regimes labeled on the top bar. (d) Doping dependence of the charge susceptibilities at $(\pi,\pi)$, the $(\pi,0)$, and the experimentally observed $(\pi/3,0)$. 
}
\end{center}
\end{figure}

To further demonstrate the above doping dependence at a lower temperature ($T=0.25t$), we switch to a $12\times4$ rectangular lattice, which is smaller than the $12\times12$ square lattice used previously. This geometry naturally breaks the $C_4$ symmetry, providing a clearer visualization of the symmetry-breaking stripe state [see Figs.~\ref{fig:HubbardRectangular}(a) and (b)]. Despite the use of slightly different model parameters, the doping dependence of the charge instability remains consistent with the underlying phases presented in Fig.~\ref{fig:Hubbard}  and Fig.~\ref{fig:appendix:HubbardCuts}. Shortly after doping, the system exhibits unidirectional stripe order along the antinodal direction. As shown in Fig.~\ref{fig:HubbardRectangular}(c), the dominant wavevector $q_x$ increases linearly with doping within the stripe regime. However, in the extremely overdoped regime beyond $30\%$ doping, the stripe charge fluctuation is then replaced by checkerboard charge fluctuations near $(\pi,\pi)$. 

Notably, the overall charge susceptibility increases with doping reaching its peak at approximately 50\% doping for both square and rectangular lattices. As elaborated in the main text, this peak emerges due to the suppression of charge fluctuations at both ends of the doping axis. At half-filling, the Mott insulating phase, driven by strong electronic correlations, minimizes double occupation and suppresses charge fluctuations. On the other hand, in the extreme hole-doped regime, where the band is nearly empty, the total charge susceptibility diminishes. Thus, the charge susceptibility reaches its maximum at the midpoint of the doping axis, around $50\%$ doping. This doping dependence is observed in $\chi(\mathbf{q})$ across all momenta, though the precise peak doping level may vary between 45\% to 55\%, depending on the specific $\mathbf{q}$ and model parameters like $U/t$. Therefore, the fact that experimentally observed susceptibility peak near 50\% suggests that strong correlations continue to play a significant role even in the extremely overdoped regime of cuprates.

From the detailed analysis in this appendix, it becomes evident that the charge susceptibility in the extremely overdoped Hubbard model is dominated by a $(\pi,\pi)$ instability on both square and rectangular lattices. While these models effectively capture the overall doping dependence of charge susceptibility, their specific momentum distributions do not align with experimental observations. This discrepancy substantiates the conclusion drawn in Sec.~\ref{sec:Hubbard} of the main text, affirming that the pure Hubbard model is insufficient to explain the experimentally observed overdoped charge order.

\section{Fermion Signs and Auto-Correlations in DQMC for the Electron-Phonon Systems}\label{app:fermionSignAutoCorr}

\begin{figure}[!t]
    \centering
    \includegraphics[width=\linewidth]{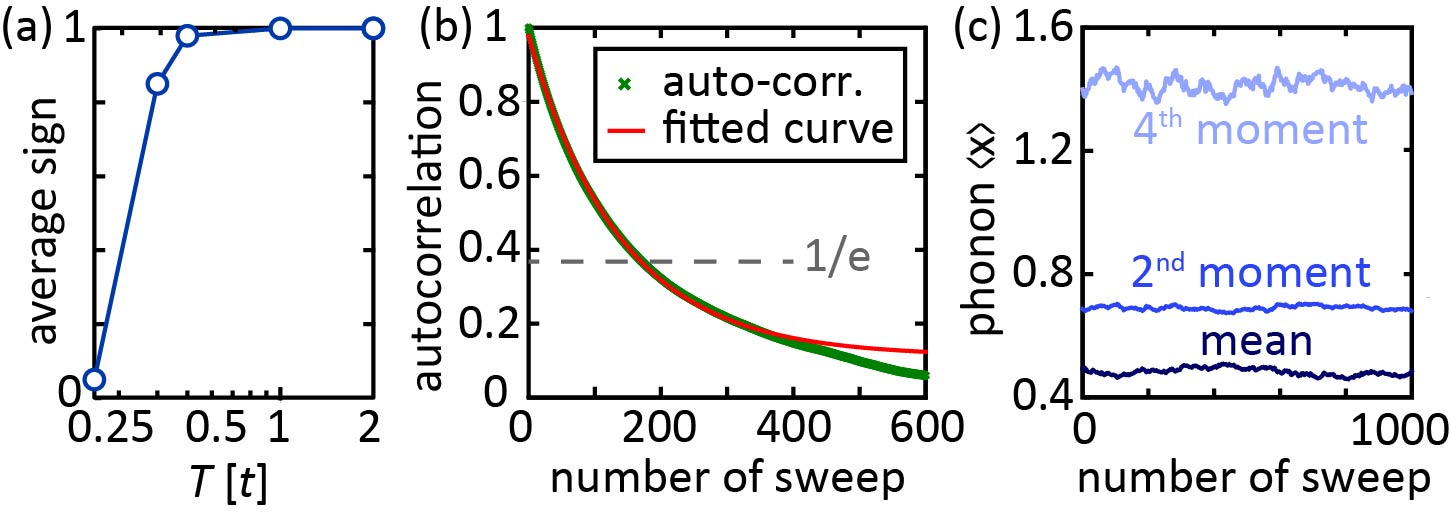}\vspace{-3mm}
    \caption{\label{fig:auto_corr}
(a) Average signs during DQMC simulations for the quarter-filling HEH model on a $12 \times 12$ square lattice across different temperatures, using parameters identical to Fig.~\ref{fig:effectiveV}. (b) Phonon autocorrelation function at $T=0.5t$, with an exponential fit shown by the red line. (c) The phonon mean displacement and second- and fourth-order momentum statistical data over 1000 sweeps from 10 processes, showing fluctuation around equilibrium value.
}    
\end{figure}

The common challenge in all DQMC simulations is the fermion sign problem, which limits the lowest temperature that can be explored in simulations of systems lacking particle-hole symmetry. A small average sign is typically indicative of very low sampling efficiency, making simulations more computationally demanding. As summarized in Fig.~\ref{fig:auto_corr}(a), for the 50\% doped HEH model in a $12 \times 12$ square lattice, with parameters , $U=8t$, $t'=-0.15t$, and $T=0.25t$ (same parameters as Fig.~\ref{fig:effectiveV} of the main text), the average sign is around 0.05, which is comparable to the average sign in the Hubbard model (0.06) under the same doping and temperature conditions. While the inclusion of phonons and relatively strong couplings does not significantly worsen the fermion sign, it substantially increases computational cost through another factor -- the autocorrelation length. In our simulations, the Hubbard-Stratonovich fields for electrons are binary, whereas the phonon fields (\textit{i.e.}, displacements) are continuous real numbers. This results in a much longer autocorrelation length for the phonon fields compared to the electronic auxiliary fields [see Fig.~\ref{fig:auto_corr}(b)], the latter usually requiring fewer than 10 sweeps. As temperatures decrease, the rapidly diverging autocorrelation length of the phonon fields makes reliable Monte Carlo sampling and achieving equilibrium states increasingly time-consuming.

To ensure the statistical reliability of the results presented, we initiate our simulations with a series of extensive preheating sweeps, starting from a random configuration where phonon displacements were distributed according to a Gaussian. Throughout this preheating phase, we monitor key statistical metrics such as the average displacement and its second and fourth-order moments to ascertain when the system reaches thermal equilibrium. For example, as shown in Fig.~\ref{fig:auto_corr}(c), these metrics converge after a sufficient number of preheating sweeps, confirming that the system has  stabilized in a thermal equilibrium state. Following this, we estimate the autocorrelation length by fitting the autocorrelation function to the exponential decay model $f(t)=\exp(-t/L)$. With the extracted autocorrelation length $L$, we conduct our measurements at intervals of $L$ sweeps to ensure statistically unbiased sampling. In the occasional scenario where the exponential fitting function does not fit the data well, we adjust the number of sweeps according to the first time point at which the autocorrelation falls below $1/e$. This strategy ensures that our DQMC simulations accurately capture the thermal equilibrium properties of the system without bias.

\section{Finite-Size Analysis of the HEH Model with Only $g$ and $g^\prime$ Couplings }\label{app:finiteSize}
\begin{figure}[!b]
    \centering
    \includegraphics[width=0.8\linewidth]{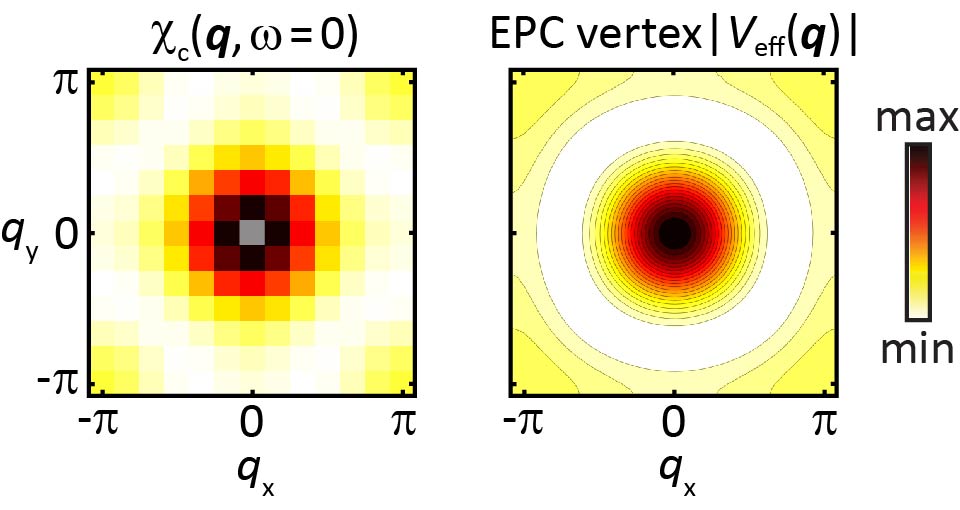}\vspace{-3mm}
\caption{\label{fig:HEHenlarged}
    Left: Charge susceptibility for the HEH model with $g=0.5t$ and $g^\prime =0.2t$, simulated on a $12\times 12$ cluster at $T=0.4t$. The remaining parameters are identical to those in Fig.~\ref{fig:HubbardExtendedHolstein}(a) of the main text. Right: Momentum-space distribution of the EPC-mediated effective interaction. 
}    
\end{figure}
Figure \ref{fig:HubbardExtendedHolstein}(a) in the main text demonstrates an increase in the charge susceptibility at the $(\pi/3,0)$ with rising $g^\prime$ on a $6\times 6$ cluster. To avoid confusion, we clarify that, under the parameters used, this peak signals the emergence of instabilities near the zone center but does not precisely determine the wavevector due to the finite system size. To further illustrate this, we repeat the simulation of Fig.~\ref{fig:HubbardExtendedHolstein}(a) on a $12\times 12$ cluster, albeit at a higher temperature ($T=0.4t$) due to the computational constraints. As shown in the left panel of Fig.~\ref{fig:HEHenlarged}, the dominant wavevector shifts closer to  $(\pi/12,0)$ instead of $(\pi/3,0)$ in this enlarged system . This shift indicates that the charge instability is expected to be dominated by a smaller wavevector near the $\Gamma$ point in the thermodynamic limit, instead of being pinned at $(\pi/3,0)$. It is also suggested by the EPC vertex distribution shown in right panel of Fig.~\ref{fig:HEHenlarged}.

Thus, this finite-size analysis shows that the HEH model with only $g$ and $g^\prime $ does not precisely stabilize the overdoped CDW at $(\pi/3,0)$, though it does induce significant fluctuations near the zone center. By comparing DQMC and RPA results in Fig.~\ref{fig:HubbardExtendedHolstein} of the main text, it becomes clear that the ordering wavevector is primarily determined by the distribution of the EPC vertex. This insight leads us to introduce additional nonlocal EPC interactions into the HEH model, ultimately achieving the precise $(\pi/3,0)$ order observed in Fig.~\ref{fig:effectiveV} of the main text, consistent with experimental observations. This progression from qualitative to quantitative analysis provides a more comprehensive explanation for the overdoped CDW phenomenon.

\section{Details about the Doping Dependence in the HEH Model}\label{app:HEH:doping}
\begin{figure}[!t]
\begin{center}
\includegraphics[width=\linewidth]{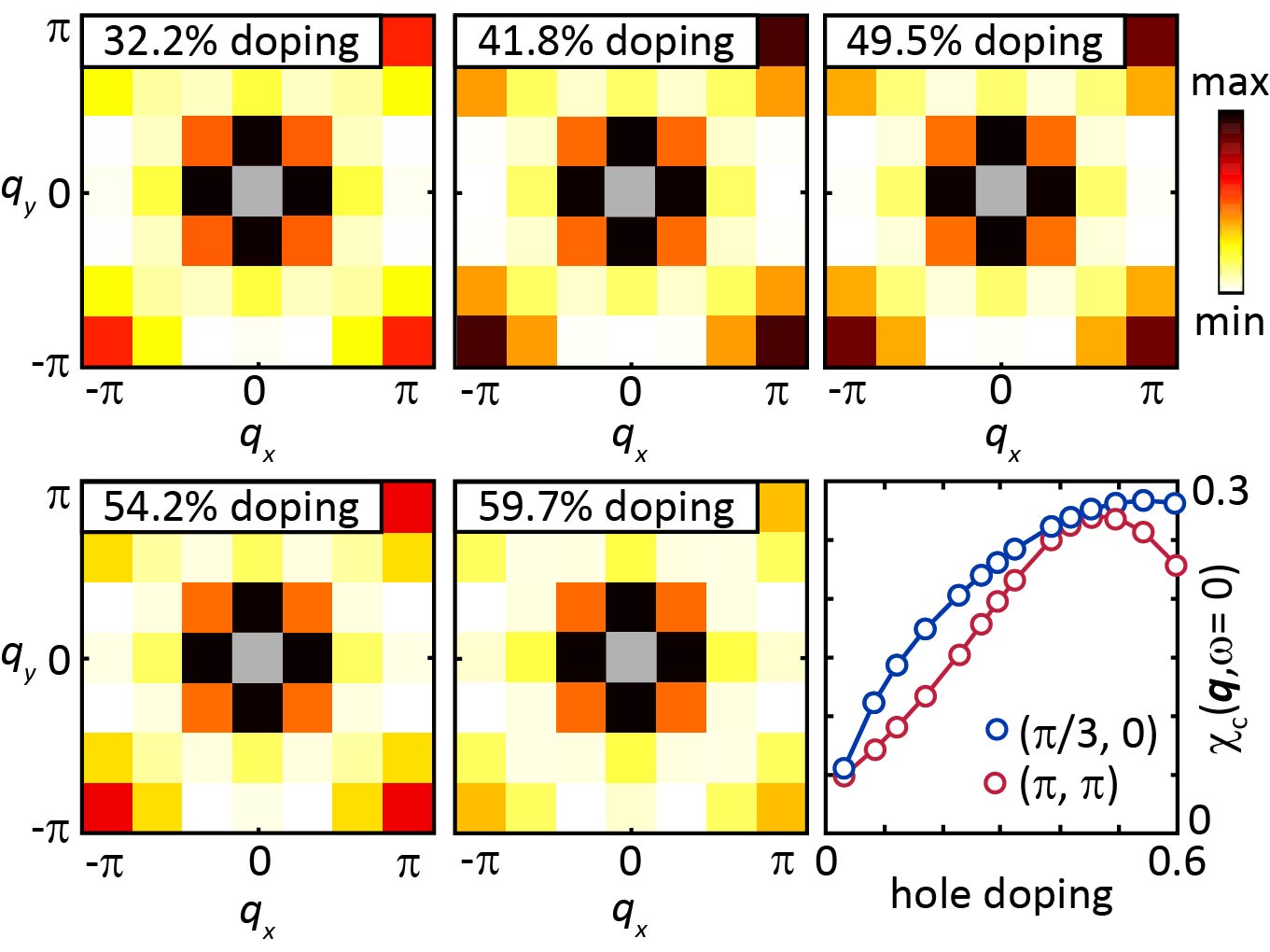}\vspace{-3mm}
\caption{\label{fig:app:HEHdopingDep}
    The doping dependence of charge susceptibilities simulated using the HEH model on a $6\times 6$ cluster with $g=0.5t$ and $g^\prime =0.3t$ at $T=0.4t$. The band structure and Hubbard interactions are identical to those in Fig.~\ref{fig:HubbardExtendedHolstein}. The last panel presents the evolution of susceptibility at $(\pi,\pi)$ and $(\pi/3,0)$. Error bars are smaller than the symbol sizes.
}
\end{center}
\end{figure}

In Fig.~\ref{fig:HubbardExtendedHolstein} of the main text, we have demonstrated that the next-nearest-neighbor EPC suppresses the checkerboard fluctuations at $(\pi,\pi)$ while enhancing charge fluctuations near the zone center at 50\% doping. To provide a more comprehensive understanding of this trend, we present the full doping dependence of charge susceptibility in Fig.~\ref{fig:app:HEHdopingDep}, using the next-nearest-neighbor coupling strength $g^\prime=0.3t$. Due to  computational complexity, these additional simulations are performed at a higher temperature $T=0.4t$. Notably, the enhancement in the zone center charge instability persists across the entire extremely doped regime. At the examined coupling strength, this zone center instability clearly dominates over the checkerboard $(\pi,\pi)$ instability. 

The maximization of charge susceptibilities near quarter doping, as discussed in Sec.~\ref{sec:Hubbard}, results from strong electronic correlations. This trend remains consistent across models, from the Hubbard model to the HEH model with EPCs, as both the coupling strength itself $g_{\mathbf{q}}$ and the EPC vertex $g_{\mathbf{q}}^2/\omega_{\rm ph}^2$ are significantly weaker than the Hubbard $U$, leaving the spin-dominant physics near half-filling unaffected. Consequently, the charge susceptibility at $(\pi,\pi)$ continues to peak at 50\% doping, whereas the zone center susceptibility, which turns to dominate in the HEH model, peaks at 55\%. While the introduction of EPCs does not alter the overall doping dependence discussed in the context of the Hubbard model, it redistributes susceptibility across different momenta. This doping dependence distinguishes our HEH model simulations, where the Hubbard $U$ remains the dominant energy scale, from previous studies on phonon-only models\,\cite{perroni2004polaron, fehske2000lattice, alexandrov1999mobile}. In those phonon-only models, the charge susceptibility consistently peaks near half filling, contrary to experimental observations. 

\section{Absence of Phase Separation in All Investigated Models}
\begin{figure}[!t]
\centering  
\includegraphics[width=\linewidth]{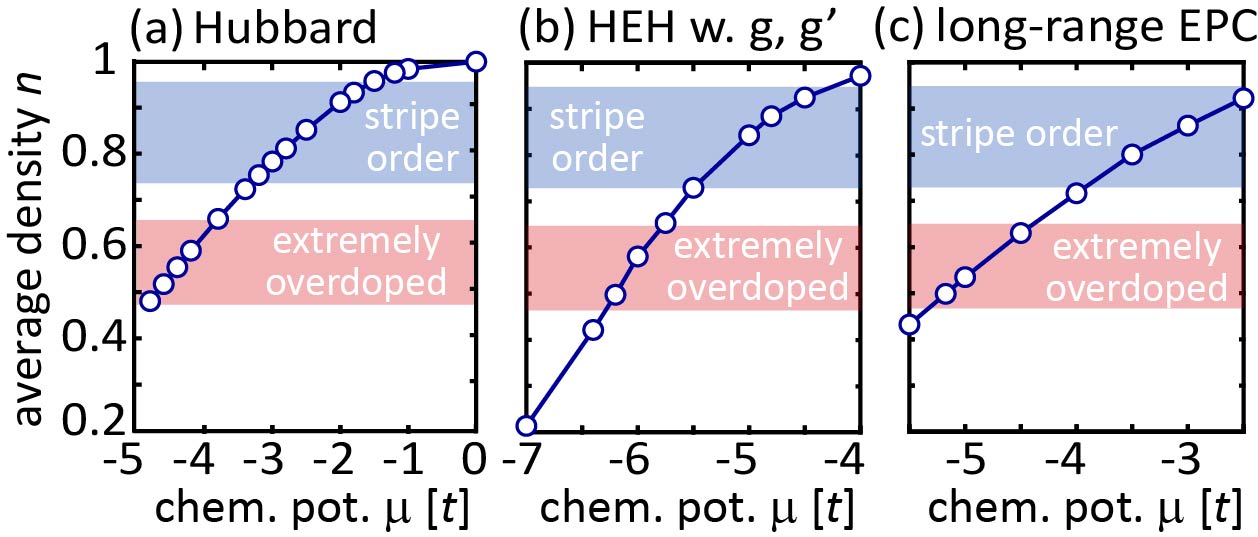}\vspace{-2mm}
\caption{\label{fig:phaseSep}
    The average density $n$ as a function of chemical potential $\mu$ for (a) the Hubbard model [parameters identical to Fig.~\ref{fig:Hubbard} of the main text], (b) the HEH model with parameters $g=0.5t$ and $g^\prime =0.3t$ [Fig.~\ref{fig:HubbardExtendedHolstein} of the main text], and (c) the HEH model with longer-range EPCs [Fig.~\ref{fig:effectiveV} of the main text], respectively. Error bars are smaller than the symbol sizes.
}    
\end{figure}

In systems where strong EPC mediates an attractive interaction, potential phase separation may occure due to the instability of particle numbers. To confirm the absence of phase separation in the models investigated here, we analyze the evolution of the average density $n$ as a function of the chemical potential $\mu$ [see Fig.~\ref{fig:phaseSep}]. A key indicator of phase separation is a discontinuity in the $n-\mu$ curve. However, for the Hubbard model, the continuous $n-\mu$ curve confirms the absence of phase separation, aligning with prior numerical studies\,\cite{jiang2019superconductivity, huang2018stripe, huang2017numerical}. The inclusion of phonons with the relevant couplings does not alter this outcome, as the $n-\mu$ relationship remains continuous across the coupling strengths considered [see Fig.~\ref{fig:phaseSep}(b)]. This observation holds true for all HEH models examined in this study, with different $g$, $g^\prime$, and other non-local interactions [see the example in Fig.~\ref{fig:phaseSep}(c)]. Therefore, phase separation is conclusively absent in the models relevant to our investigation.

The stability observed in our models largely stems from the strong Hubbard interaction $U=8t$. This repulsive interaction dominates the energy landscape of the system and effectively prevents the phase separation that might be induced by attractive phonon-mediated interactions. It is generally recognized that for phase separation to occur, the attractive interaction must reach approximately half the magnitude of the Hubbard $U$\,\cite{qu2022spin,chen2023superconducting}. Our study confirms that, while the phonon-mediated attractive $V$ is significant, it does not exceed this critical threshold set by $U$. This observation further highlights the importance of incorporating the strong electronic correlation in the study of phonons.

\bibliography{references}

\end{document}